\documentclass[onecolumn]{aastex63}

\usepackage{amsmath}
\usepackage{multirow}
\usepackage{textcomp}
\usepackage{color, colortbl}
\usepackage{enumitem}

\newcommand{\Rs}{R$_\odot$}
\newcommand{\Bs}{B$_\odot$}

\newcommand{\mic}{\textmu m}
\newcommand{\revis}[1] {#1}
\newcommand{\revisp}[1] {#1}

\def \specrad {photons s$^{-1}$ cm$^{-2}$ sr$^{-1}$ \AA$^{-1}$}

\definecolor{LightCyan}{rgb}{0.88,1,1}

\shorttitle{2019 AIR-Spec Eclipse Observations}
\shortauthors{Samra et al.}

\begin{document}

\title{New Observations of the IR Emission Corona from the July 2, 2019 Eclipse Flight of the \\Airborne Infrared Spectrometer}
\correspondingauthor{Jenna Samra}
\email{jsamra@cfa.harvard.edu}

\author{Jenna E. Samra}
\affiliation{Smithsonian Astrophysical Observatory, 60 Garden Street, Cambridge, MA, USA}

\author{Chad A. Madsen}
\affiliation{Smithsonian Astrophysical Observatory, 60 Garden Street, Cambridge, MA, USA}

\author{Peter Cheimets}
\affiliation{Smithsonian Astrophysical Observatory, 60 Garden Street, Cambridge, MA, USA}

\author{Edward E. DeLuca}
\affiliation{Smithsonian Astrophysical Observatory, 60 Garden Street, Cambridge, MA, USA}

\author{Leon Golub}
\affiliation{Smithsonian Astrophysical Observatory, 60 Garden Street, Cambridge, MA, USA}

\author{Vanessa Marquez}
\affiliation{Smithsonian Astrophysical Observatory, 60 Garden Street, Cambridge, MA, USA}

\author{Naylynn Ta\~n\'on Reyes}
\affiliation{Smith College, Northampton, MA, USA}

\begin{abstract}
	The Airborne Infrared Spectrometer (AIR-Spec) was commissioned during the 2017 total solar eclipse, when it observed five infrared coronal emission lines from a Gulfstream V (GV) research jet owned by the National Science Foundation (NSF) and operated by the National Center for Atmospheric Research (NCAR). The second AIR-Spec research flight took place during the July 2, 2019 total solar eclipse across the south Pacific. The 2019 eclipse flight resulted in seven minutes of observations, during which the instrument measured all four of its target emission lines: \ion{S}{11}~1.393~\mic, \ion{Si}{10}~1.431~\mic, \ion{S}{11}~1.921~\mic, and \ion{Fe}{9}~2.853~\mic. The 1.393~\mic\  line was detected for the first time\revis{, and probable first detections were made of \ion{Si}{11}~1.934~\mic\ and \ion{Fe}{10}~1.947~\mic.} The 2017 AIR-Spec detection of \ion{Fe}{9} was confirmed and the first observations were made of the \ion{Fe}{9} line intensity as a function of solar radius. Telluric absorption features were used to calibrate the wavelength mapping, instrumental broadening, and throughput of the instrument. AIR-Spec underwent significant upgrades in preparation for the 2019 eclipse observation. The thermal background was reduced by a factor of 30, providing a 5.5x improvement in signal-to-noise ratio, and the post-processed pointing stability was improved by a factor of five to $<$10 arcsec rms. In addition, two imaging artifacts were identified and resolved, improving the spectral resolution and making the 2019 data easier to interpret. \\
	\phantom{0}\\
	\phantom{0}\\
	\phantom{0}
\end{abstract}



\section{Introduction}
\label{sec:intro}
Infrared (IR) solar coronal emission lines \citep{Judge1998,DelZanna2018} are currently of significant interest due to their advantages for coronal magnetic field measurements and coronal plasma characterization at large solar radii. These magnetic dipole transitions encode the magnetic field through both the ``weak-field'' Zeeman effect and the ``strong field limit'' of the Hanle effect \citep{Casini1999,Lin2000,Judge2001}.  Unlike their ultraviolet (UV) and X-ray counterparts, visible and IR lines are excited in part by resonant scattering of photospheric radiation and fall off in intensity more gradually with radius \citep{Judge1998,Habbal2011}. Near and mid-IR lines are particularly promising because they balance Zeeman sensitivity, which increases as $\lambda^2$, with line intensity, which generally decreases with wavelength \citep{Judge2001}. This wavelength region also optimizes the instrumental tradeoff between thermal emission, which increases nearly exponentially with wavelength, and Rayleigh scattering, which decreases as $\lambda^4$ \citep{Judge2001}. 

\revis{The IR emission corona was first and is most frequently observed below 1.1 \mic, where silicon detectors can be used and telluric absorption and room-temperature thermal emission are negligible. In 1959, \citet{Firor1962} detected the forbidden \ion{Fe}{13} 1.075 \mic/1.080 \mic\ line pair using a coronagraph under daylight conditions. This line pair is exceptionally intense, making it accessible even to coronagraphic IR instrumentation with significant scattering background. The pair is an excellent diagnostic for electron density, as \cite{Singh2002} demonstrated using the 25~cm coronagraph at Norikura Observatory.  Later, advances in IR instrumentation and detectors made it possible to observe coronal lines with longer wavelengths. The 1.43 \mic\ line of \ion{Si}{10} was first observed during an eclipse \cite{Mangus1966,Munch1967}, but \cite{PennKuhn1994} later made precise measurements of its wavelength, FWHM, and intensity using a National Solar Observatory coronagraph at Sacramento Peak. \cite{Judge2002} measured the 3.93 \mic\ line of \ion{Si}{9} for the first time using the McMath-Pierce telescope on Kitt Peak.  Coronagraphs have been used to measure the strongest IR lines not just in intensity, but also in linear polarization.} The Coronal Multi-Channel Polarimeter \citep[CoMP,][]{Tomczyk2008}, installed at the Mauna Loa Solar Observatory (MLSO) in 2010, measures the full polarization state in the 1.075/1.080 \mic\ \ion{Fe}{13}  line pair. The MLSO \revis{has recently deployed} an upgrade to CoMP that expands wavelength coverage to include visible and additional near-IR lines \citep{Landi2016,Tomczyk2019}. \revis{The first coronagraphic linear polarization measurements of the 1.43 \mic\ \ion{Si}{10} line were made by \cite{Dima2019} using the Scatter-free Observatory for Limb, Active Regions and Coronae (SOLARC, \citealp{Kuhn2003}) on Haleakal\={a}.}

Significant advances are anticipated from the Daniel K. Inouye Solar Telescope (DKIST, \citealp{Tritschler2016,Rimmele2020}). Its unique coronagraphic and IR capabilities should enable \revis{measurements of coronal emission line polarization and} magnetic fields with high spatial resolution and high sensitivity \citep{Rast2020}. The Cryogenic Near Infrared Spectropolarimeter (Cryo-NIRSP, \citealp{Fehlmann2016}), a DKIST focal plane instrument, will make the first regular measurements of mid-IR coronal lines with wavelengths as long as 5 \mic. In the near future, the Visible Emission Line Coronagraph (VELC), part of the Aditya mission to L1, will observe two visible emission lines and make spectropolarimetric measurements of the 1.07 \mic\ Fe XIII line \citep{Singh2019}.

Total solar eclipses offer a unique opportunity to make visible and IR observations of the corona without a coronagraph.  Eclipse observations of visible emission lines provided the first indication of the corona's high temperature when \citet{Edlen1943} showed that these lines arose from transitions in highly ionized species. \revis{Later on, \citet{Blackwell1952} made the first IR observations of a total solar eclipse, supporting the existence of the F-corona via radiometric measurements of IR excesses. \citet{Taylor1964} expanded upon the work of \citet{Blackwell1952} by implementing wide-band interference filters to isolate distinct radiometric passbands in an attempt to verify contemporaneous structural models of the corona. These early radiometric studies informed future IR imaging campaigns under total solar eclipse conditions, notably a string of observations to detect coronal dust originating from solar-cometary interactions \citep{Mizutani1984,Lamy1992,Hodapp1992,Tollestrup1994}. During the 2017 eclipse, \cite{Judge2019} surveyed the visible and IR solar atmosphere out to 5.5 \mic\ using a suite of imagers, spectrometers, and a polarimeter deployed from Casper Mountain, Wyoming.}

\revis{Eclipse observations have also enriched the study of IR coronal line emission. \citet{Kurt1962} detected the \ion{Fe}{13} 1.075 \mic/1.080 \mic\ line pair during a total solar eclipse in 1961. This spurred considerable interest in observing the pair during subsequent eclipse events \citep[e.g.][]{Eddy1967,Byard1971,Pasachoff1976,Penn1994,Bao2009}. Steady advancements in detector sensitivity and cryogenic techniques allowed several eclipse observing campaigns to significantly expand the known coronal IR spectrum, including observations resulting in detections of \ion{S}{9} 1.25 \mic, \ion{Fe}{14} 1.27 \mic, \ion{Si}{10} 1.43 \mic, \ion{S}{11} 1.92 \mic, \ion{Al}{10} 2.75 \mic, \ion{Mg}{8} 3.03 \mic, and \ion{Si}{9} 3.93 \mic\ \citep{Mangus1966,Munch1967,Olsen1971,Kastner1993,Kuhn1996,Kuhn1999,Dima2018}}. Recently, Habbal and collaborators \citep{Boe2018Freeze,Boe2020topology,Boe2020CME,Habbal2021} have shown how multi-wavelength observations can be used to estimate the freeze-in height of ions, measure the plasma properties of CMEs, and connect coronal fields into the low heliosphere. These papers highlight the rich diagnostic capabilities of visible and near IR eclipse observations.

\revis{Telluric absorption remains a significant hurdle in further surveying near and mid-IR coronal emission during total solar eclipses. This drives the need to extend observing campaigns into higher altitudes above most of the IR-opaque material in Earth's atmosphere. Opportunistic studies from high-altitude ground-based observatories have proved fruitful \citep[e.g.][]{Kuhn1994,Judge2002}, but such campaigns rely on ideal weather conditions and geographic accessibility, and even the highest ground-based observatories only reach altitudes of about 4 km. Airborne observatories are capable of reaching much higher altitudes and can capture eclipse totality along any point of the eclipse path. \citet{Kurt1962} was the first to attempt airborne IR eclipse observations from an aircraft, detecting the \ion{Fe}{13} 1.075 \mic/1.080 \mic\ line pair from an altitude of 10 km. \citet{Mangus1966} flew a 1--3.5 \mic\ Fourier transform spectrometer in the 1965 eclipse and made the first probable detection of \ion{Si}{10} 1.43 \mic.   This detection was shortly confirmed by the 1966 airborne observation of \citet{Munch1967}, who also measured \ion{Mg}{8} 3.03 \mic. Later, \citet{Byard1971} observed aboard a NASA CV990 at an altitude of 12 km with the intent to produce precise intensity measurements of the \ion{Fe}{13} pair. \citet{Kuhn1999} flew a cryogenically cooled instrument assembly aboard an open C130 aircraft during the 1998 February 22 total solar eclipse.  The assembly contained broadband and narrowband IR imagers as well as an IR spectrograph which successfully detected Si IX 3.93 \mic\ and measured the radial brightness profile of the F-corona \citep{Ohgaito2002}. However, the experiment suffered greatly from poor pointing stability ($>$100 arcsec rms) due to high-frequency vibrations from the aircraft, limiting the spatial resolution of the imagers and spectrograph as well as preventing sampling of small-scale coronal structures. More recently, \cite{Caspi2020} observed the 3--5 \mic\ continuum from two NASA WB-57 aircraft during the 2017 eclipse.}

The Airborne Infrared Spectrometer (AIR-Spec) is a new \revis{airborne cryogenic grating spectrometer} \citep{Samra2021} funded by the NSF and developed by Smithsonian Astrophysical Observatory (SAO) to characterize near and mid-IR emission lines as a function of solar conditions and radius. The instrument was commissioned during the 2017 eclipse \revis{and an upgraded version observed the 2019 eclipse over the South Pacific. Its sensitivity and spatial resolution distinguish AIR-Spec from previous airborne eclipse instruments. In two eclipse flights, AIR-Spec made the first detection of \ion{Fe}{9}~2.853~\mic\ and \ion{S}{11}~1.393~\mic\ and the first probable detections of the weak lines \ion{Si}{11}~1.934~\mic\ and \ion{Fe}{10}~1.947~\mic. It measured emission line intensity up to 1~\Rs\ from the solar limb, providing coronal temperature and density estimates (\citealp{Madsen2019} and a paper in preparation) and an understanding of which lines are promising for future ground-based, airborne, and space-based spectropolarmetic observation.}

This paper describes the 2019 instrument, campaign, data processing, and observations.  \revis{Section \ref{sec:airspec} gives an overview of the instrument and describes the improvements made ahead of the 2019 observation.} Section \ref{sec:campaign} details the flight campaign.  Section \ref{sec:analysis} describes the analysis of the spectra, including preparation of the data, the spectral fitting routine, and the calibration products derived from the fit. Section \ref{sec:results} discusses the scientific results and compares the AIR-Spec measurements with models. Section \ref{sec:data} briefly describes the available data.

\section{\revis{The Airborne Infrared Spectrometer}}
\label{sec:airspec}

\revis{AIR-Spec consists of an image stabilization system, a telescope, a broadband visible-light context camera, and a cryogenic grating spectrometer that observes in three (previously four) IR passbands over a 1.5~\Rs\ field of view.  The passbands are measured in two overlapping diffraction orders. Order-sorting filters are omitted in order to collect coronal observations in both overlapping passbands for the entire duration of totality. The spectrometer optics are housed in a vacuum chamber and cooled to $<$150 K with liquid nitrogen. The spectrometer focuses light onto an InSb detector cooled to 50 K (previously 59 K) by a closed-cycle cooler. The front end of the camera includes an aperture stop and bandpass filter to reject photons from outside the field of view and passband, respectively.  Both the stop and the filter are cooled to the detector temperature of 50 K. A comprehensive description of the instrument is given in the AIR-Spec instrument paper \citep{Samra2021}.}

\begin{figure}[h]
		\centering
		\includegraphics[angle=0,width=0.85\linewidth]{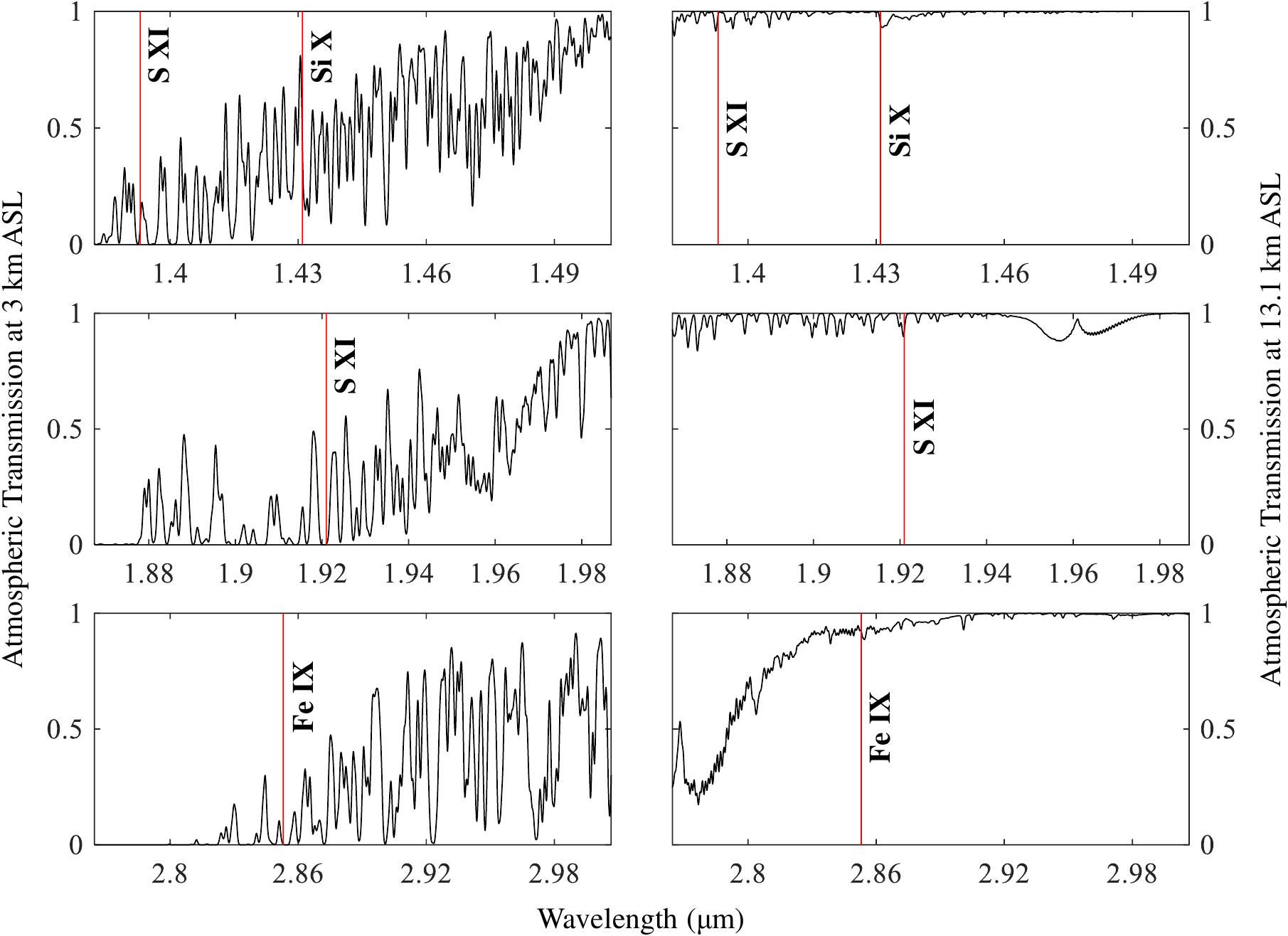}
		\caption{\revisp{ATRAN-modeled atmospheric transmission \citep{Lord1992} in the three AIR-Spec passbands at 3 km (left) and 13.1 km (right) above sea level, after convolution with the instrument line-spread function.}}
		\label{fig:atm-model}
	\end{figure}

\revis{AIR-Spec flies on the NSF/NCAR Gulfstream V (GV) at an altitude of at least 12 km, giving it access to emission lines that are almost completely absorbed at the ground. Figure \ref{fig:atm-model} compares the atmospheric transmission in the 2019 AIR-Spec passbands at 3 km and 13.1 km above sea level, the altitudes of DKIST and the 2019 AIR-Spec observation. The transmission spectra were estimated using the web-based ATRAN model at \url{https://atran.arc.nasa.gov} \citep{Lord1992}, assuming the 40.3$^\circ$ zenith angle of the 2019 eclipse, and smoothed to the AIR-Spec spectral resolution. Three of the four AIR-Spec target lines are almost completely absorbed at 3 km but have greater than 90\% transmission at 13.1 km altitude.}

\revis{Unlike IR coronagraphs, AIR-Spec is not sensitive to scattered light. The total eclipse ensures that no photospheric light reaches the aircraft or instrument, and the coronal light scattered by the atmosphere, aircraft window, and instrument is negligible compared to the instrument thermal background.  The total integrated scatter $TIS$ from one surface of the aircraft viewport can be estimated from the rms surface roughness $\sigma$ as}

\begin{equation}
TIS=\left[\frac{2\pi\sigma\left(n_{sapphire}-n_{air}\right)}{\lambda}\right]^2 \textrm{\citep[Eq. 4.14]{Fest2013}},
\label{eq:scatter}
\end{equation}

\noindent\revis{where $\lambda$ is the wavelength of the light and $n_{sapphire}$, $n_{air}$ are the refractive indices of sapphire and air at $\lambda$.  Assuming a wavelength of 1.39 \mic\ (the shortest AIR-Spec line wavelength and worst case for scatter) and a 30~\AA\ rms surface roughness (routinely achievable with sapphire), we estimate the per-surface $TIS$ to be $1\times10^{-4}$. Together, the four surfaces of the double-paned viewport scatter with efficiency $4\times10^{-4}$. The scattered light is over three orders of magnitude lower than the coronal signal.}

\revisp{During its 2017 eclipse observation, AIR-Spec observed five emission lines between 1.4 and 4 \mic\ (Table \ref{tab:target-lines}). The 2.853~\mic\ Fe IX line was detected for the first time \citep{Samra2018}, and coordinated observations with the Hinode Extreme Ultraviolet Imaging Spectrometer \citep[EIS,][]{Culhane2007} were used to characterize plasma temperature and density \citep{Madsen2019}.}

\begin{deluxetable}{ccccc}
	\tablecaption{\revisp{AIR-Spec target lines.}}
	\label{tab:target-lines}
	\tablehead{\colhead{\revisp{Ion}} & \colhead{\revisp{Temp}} & \colhead{\revisp{Wavelength}} & \revisp{2017} & \revisp{2019} }
	\startdata
	\revisp{\ion{S}{11}} & \revisp{1.8 MK}& \revisp{1.393 \mic}& &\revisp{X}\\
	\revisp{\ion{Si}{10}} & \revisp{1.4 MK}& \revisp{1.431 \mic}&\revisp{X} &\revisp{X}\\
	\revisp{\ion{S}{11}} & \revisp{1.8 MK}& \revisp{1.921 \mic}&\revisp{X} &\revisp{X}\\
	\revisp{\ion{Fe}{9}} & \revisp{0.8 MK}& \revisp{2.853 \mic}&\revisp{X} &\revisp{X}\\
	\revisp{\ion{Mg}{8}} & \revisp{0.8 MK} & \revisp{3.028 \mic}&\revisp{X}& \\
	\revisp{\ion{Si}{9}} & \revisp{1.1 MK} & \revisp{3.935 \mic}&\revisp{X}&
	\enddata
\end{deluxetable}

\revisp{The second AIR-Spec research flight took place during the July 2, 2019 total solar eclipse across the South Pacific.  The passband was adjusted to measure the 1.393 \mic\ \ion{S}{11} line, half of a density-sensitive line pair, at the expense of the \ion{Mg}{8} line. The long-wavelength cutoff was shifted from 4 \mic\ to 3 \mic\ in order to reduce the instrument's thermal background, resulting in the loss of the \ion{Si}{9} line.  Sensitivity (Section \ref{sec:therm-bg}) and image stability (Section \ref{sec:img-stab}) were improved dramatically, and the instrument was modified to eliminate two  artifacts that appeared in the 2017 data (Section \ref{sec:artifacts}). In addition, in-flight operations and data post-processing were improved based on experience in 2017.}

\subsection{Thermal Background Reduction}
\label{sec:therm-bg}
\begin{figure}[h]
	\centering
	\includegraphics[angle=0,width=0.82\linewidth]{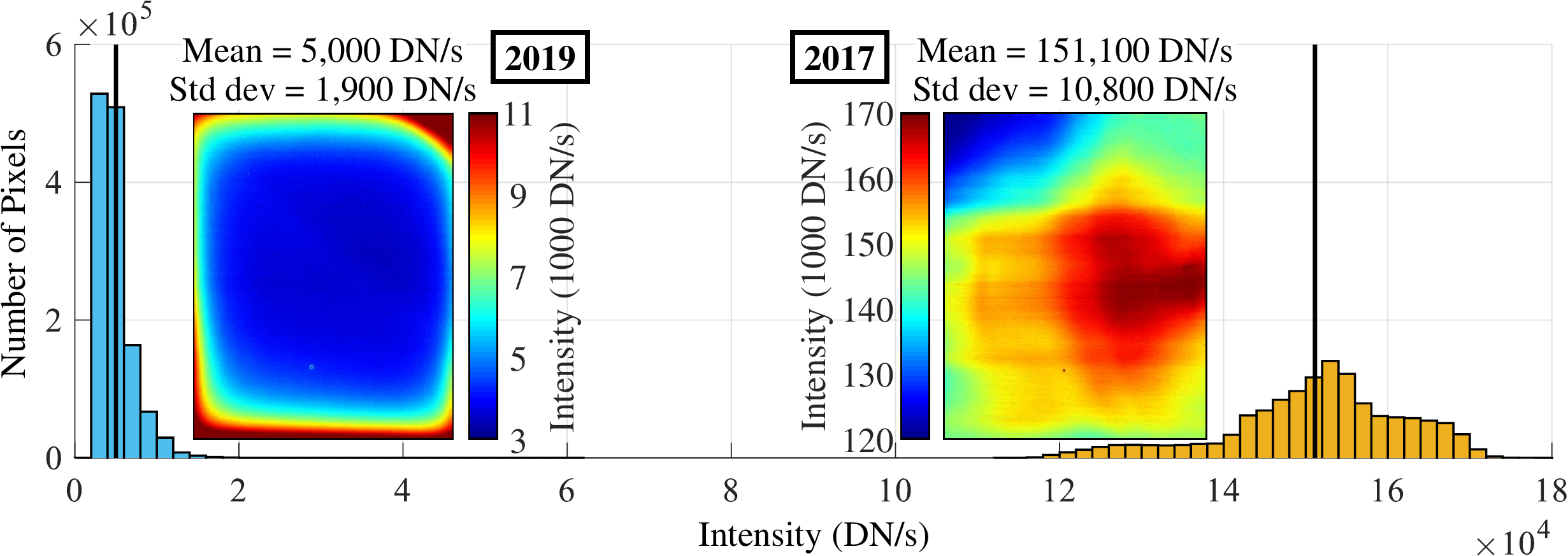}
	\caption{Thermal background improvement from 2017 to 2019. The underlying graph compares histograms of the 2017 and 2019 dark frames. The dark frames themselves are shown in the overlaid images. The 2019 background is 30 times lower on average and has significantly less spatial structure.  }
	\label{fig:therm-bg}
\end{figure}

Noise in the AIR-Spec measurements is dominated by the instrument background.  During the 2017 observation, the combination of thermal light leaks and dark current resulted in an instrument background of over 150,000 DN/s, over 60 times higher than the peak of the strongest emission line (\ion{Si}{10} 1.43 \mic) in the inner corona.  Before the 2019 observation, we tested a number of different possible sources for the light leaks and finally traced them to the camera housing itself. The camera was returned to the manufacturer for an upgrade to the front end, including an improved light baffle and a narrower bandpass filter that cut out the thermal background in the 3--4~\mic\ region at the expense of the \ion{Si}{9} emission line. The focal plane temperature was decreased \revis{from 59 K to 50 K} in order to reduce the dark current. To remove the piezoelectric signal from the chiller piston, the exposure time was set as close as possible to the reciprocal of the frame rate \citep[Sec. 3.2.1 and Eq. 1]{Samra2021}. The final result was a  background level 30 times lower than in 2017 and with significantly less spatial structure and temporal variation. This improvement, shown in Figure \ref{fig:therm-bg}, resulted in more than five times the signal-to-noise ratio (SNR) and significantly less error in the background subtraction (Section~\ref{sec:dark-sub}). \revis{Even with these improvements, the thermal background is on the order of the coronal signal near the lunar limb. The coronal light scattered by the aircraft window ($4\times10^{-4}$ scattering efficiency) is negligible in comparison.}

\subsection{Image Stability Improvement}
\label{sec:img-stab}
While preparing for the 2017 eclipse, we made the strategic decision to guide the image stabilization system based entirely on measurements of the airplane's location and orientation. We avoided implementing a control system that would actively stabilize the image based on real-time frames from the slit-jaw camera because we had no way to test that system under eclipse conditions. During the AIR-Spec rebuild, we used slit-jaw images from the 2017 observation to design a closed-loop image stabilization system \citep{Vira2018,Menzel2018}. The error signal from the slit-jaw camera is computed by fitting a circle to the lunar limb \citep{Gander1994}. The camera, a Prosilica GX1050 from Allied Vision Technologies, provides a 3.4 \Rs\ field of view sampled by 3.14 arcsec pixels. The field of view captures nearly half of the lunar limb when the slit is centered over the corona, and the pixel size provides up to 1000 pixels along the limb. The 50 Hz frame rate is sufficient for the control system to remove low-frequency drift from the image.   The open-loop system from 2017 \citep[Sec. 2.2.2]{Fedeler2017,Samra2021} corrects the high-frequency jitter using gyroscope measurements at 500 Hz.

\begin{figure}
\gridline{\fig{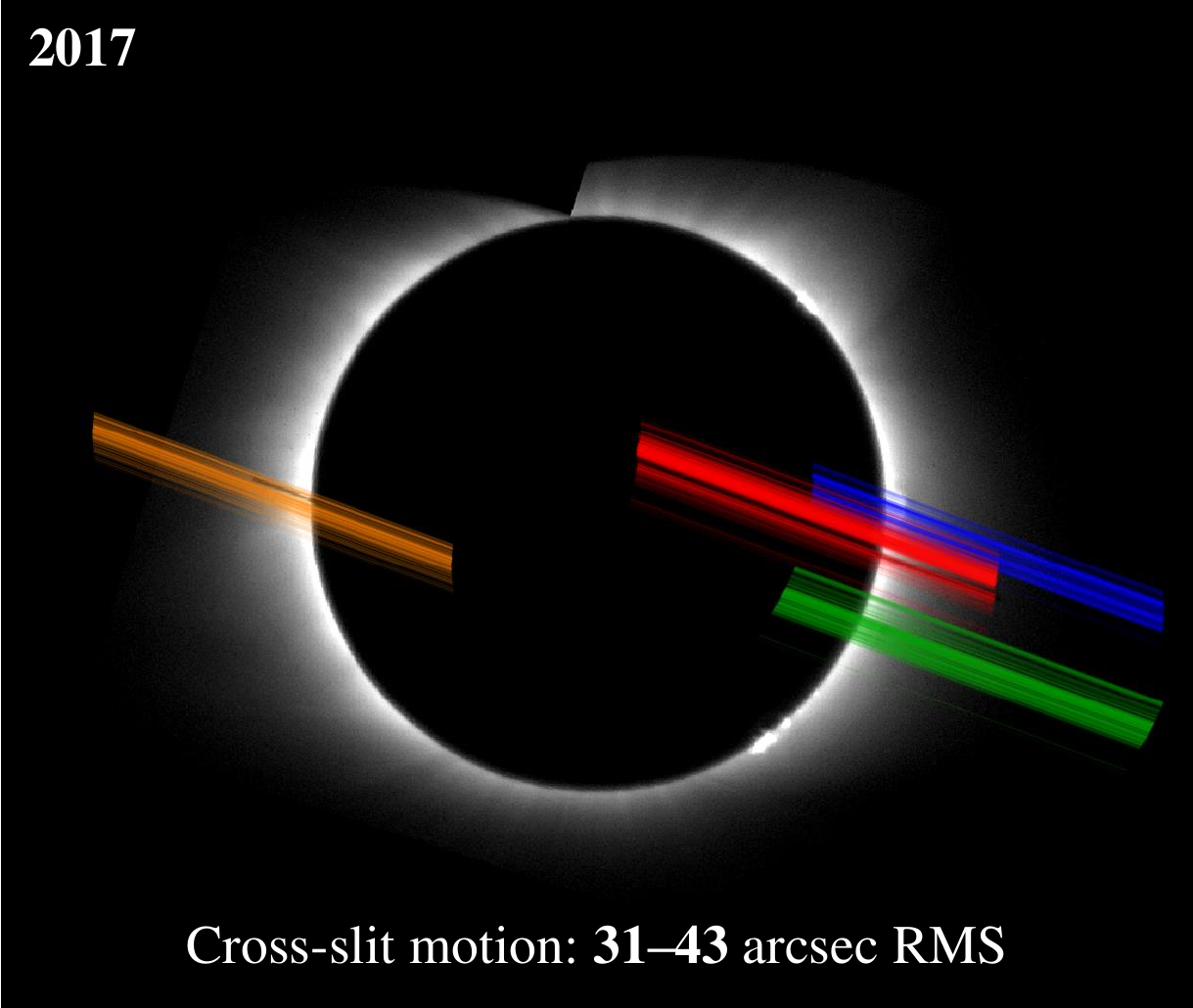}{0.4\textwidth}{(a)}
          \fig{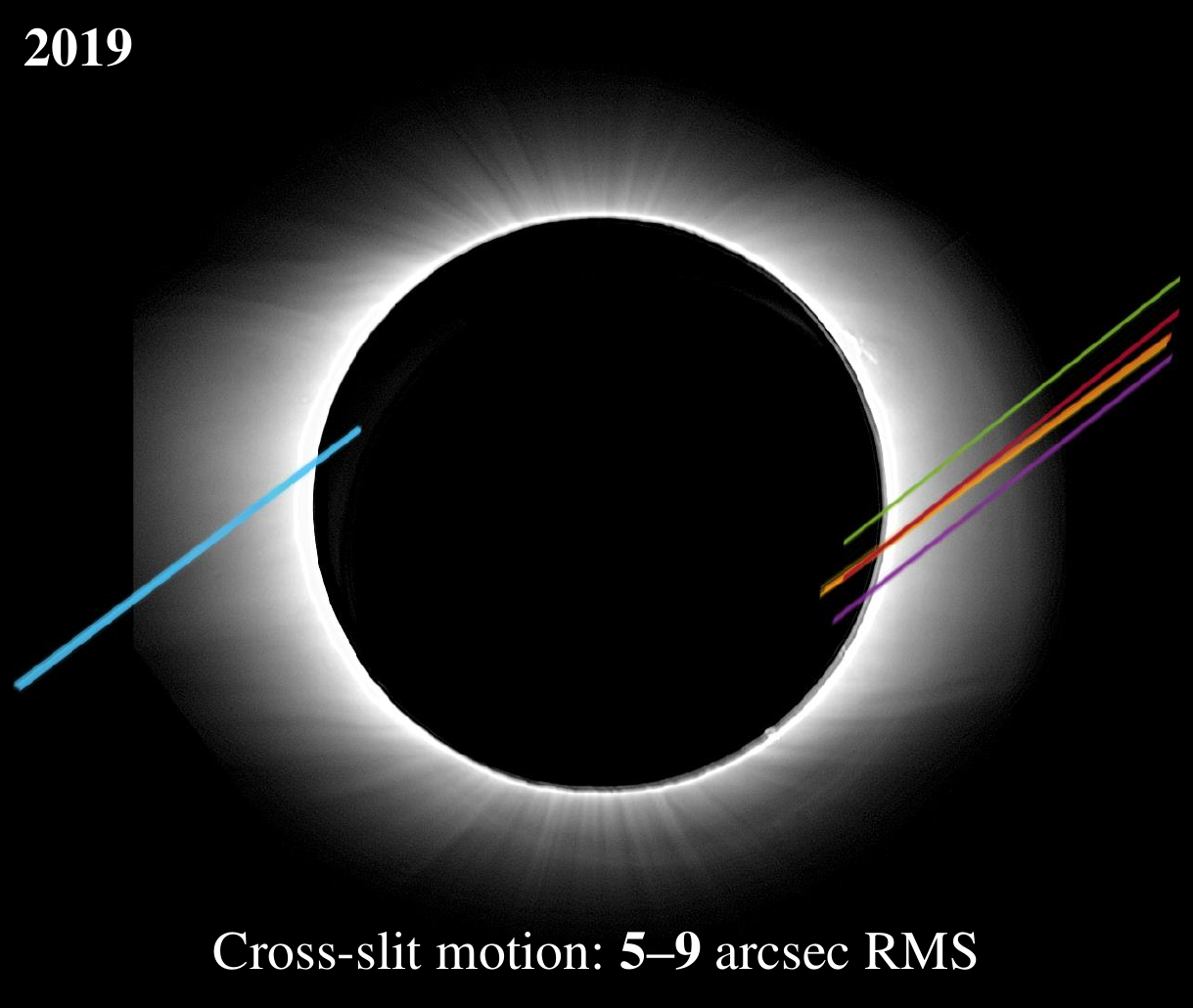}{0.4\textwidth}{(b)}}
\caption{Image stability improvement from 2017 (a) to 2019 (b).}
\label{fig:img-stab}
\end{figure}

\begin{figure}[h]
	\centering
	\includegraphics[angle=0,width=0.88\linewidth]{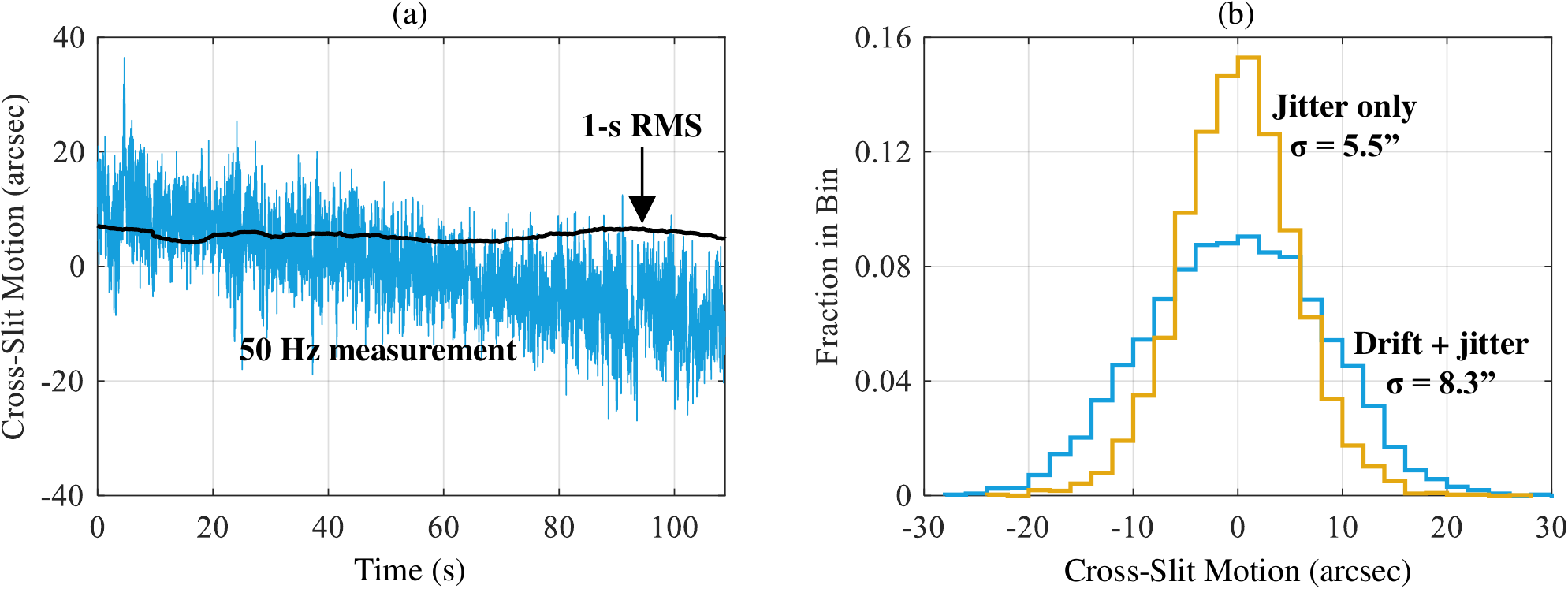}
	\caption{Time series (a) and histograms (b) of image motion perpendicular to the slit from a 109-s observation at the east limb. Image motion is measured using the slit-jaw camera.  The secular drift in (a) arises from the fact that the instrument tracks the lunar limb instead of the solar limb. A linear trend was removed from the ``jitter only'' histogram in (b).}
	\label{fig:stability}
\end{figure}

This combination of feedforward and feedback control suppresses the image motion to a greater degree than the feedforward-only system used in 2017. Figure \ref{fig:img-stab} shows the qualitative improvement in image stability from 2017 to 2019. After co-aligning frames in the direction parallel to the slit, the remaining cross-slit spread at each coronal position is 5--6 times lower in 2019. Figure \ref{fig:stability} quantifies the jitter in the east limb observation. Over a 1-s exposure time, the root-mean-square (rms) image motion is 4--7 arcsec. Once the drift of the lunar limb with respect to the solar limb is removed, the entire 109-s observation has a cross-slit standard deviation of only 5.5 arcsec. 

\revis{In addition to causing spatial image jitter, temporal variations in the aircraft yaw, pitch, and roll couple into the spectrometer and shift its pixel-to-wavelength mapping. The wavelength shift in spectral pixels is much smaller than the image motion in spatial pixels, but unlike image jitter, it is not correctable in real time. Fortunately, the $\sim$10 second period of the spectral motion is much longer than the 0.25 second exposure time, so the spectral resolution is not affected. More details on this effect are given in Section \ref{sec:spec-drift-dist}}.

\subsection{Spectral Artifact Mitigation}
\label{sec:artifacts}
As part of the 2019 rebuild, we made two changes to the spectrometer to remove optical artifacts that appeared in the 2017 data. In the process of re-aligning AIR-Spec, we discovered that internal reflections in the sapphire slit-jaw substrate were producing multiple images of the slit on the detector. This resulted in a false detection in 2017 \citep{Samra2019}. We fixed this by adding a metal plate behind the slit-jaw to prevent these reflections from propagating through the spectrometer. Double peaks were observed in some of the emission lines measured in 2017 \citep[Sec. 3.3.1]{Samra2021}. We traced this artifact to vibration in the spectrometer mirrors, which were not well-coupled to their tilt stages. Once we constrained this mirror, we saw no sign of the artifact in 2019.

\section{2019 Eclipse Campaign}
\label{sec:campaign}

\begin{figure}[h]
	\centering
	\includegraphics[angle=0,width=0.4\linewidth]{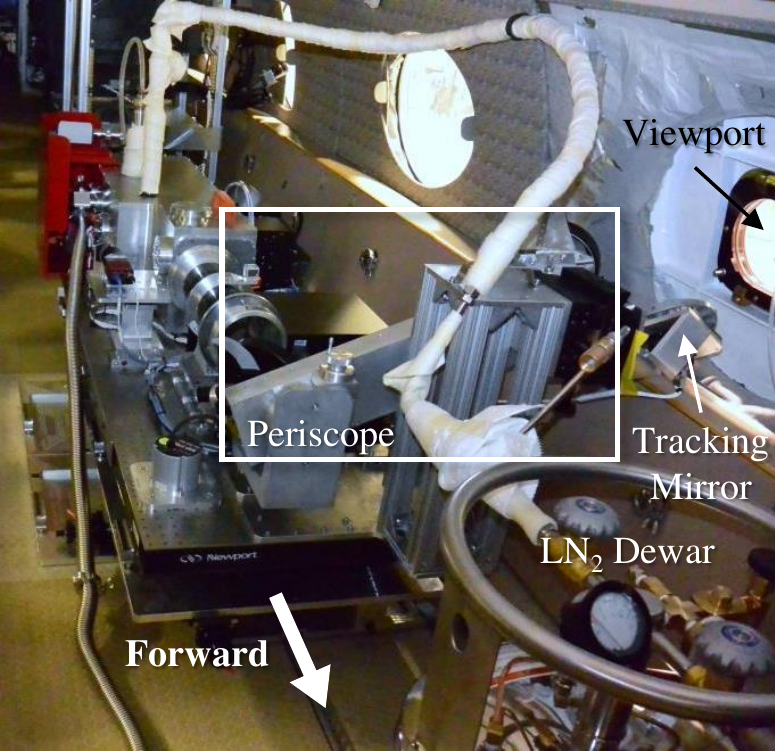}
	\caption{AIR-Spec in the GV during the 2019 eclipse campaign.}
	\label{fig:airspec}
\end{figure}

The 2019 eclipse took place in winter in the Southern Hemisphere. \revis{During the total eclipse observation, the aircraft altitude was 13.1 km (43,000 ft) above sea level, about 1.2~km lower than in 2017.} The Sun was relatively low in the sky compared to the 2017 eclipse and north of the eclipse path instead of south. The eclipse moved from west to east across the Pacific Ocean.   As in 2017, AIR-Spec was installed on the port side of the GV cabin and observed the eclipse out of a double-paned $150\times220$ mm sapphire viewport, but this time the viewport was installed in one of the port-side passenger windows. In order to address the $50^{\circ}$ solar elevation angle, the tracking mirror was placed very close to the viewport. A periscope, not present in 2017, relayed light from the tracking mirror into the telescope. Liquid nitrogen (LN$_2$) was carried on-board due to the long transit time from the base of operations in Lima, Peru to the eclipse observation point over the south Pacific. Figure \ref{fig:airspec} shows AIR-Spec mounted in the GV with the viewport, periscope, tracking mirror, and LN$_2$ dewar labeled.

\subsection{Flight and Operations}
\label{sec:flight-ops}
AIR-Spec requires at least eight hours of cooling to achieve a stable thermal environment and a vacuum level that can be sustained for hours without pumping. We began pumping and chilling AIR-Spec at 1000 UT on July 2, more than nine hours ahead of the eclipse observation. About two hours later, we prepared for takeoff by disconnecting the vacuum pump and ground-based LN$_2$ tank, topping up the spectrometer with a hand pour, and connecting AIR-Spec to the on-board dewar shown in Figure \ref{fig:airspec}.  From that point on, the spectrometer pumped itself internally with a cryo-pump, and we sustained the temperature with periodic (roughly hourly) infusions of the limited LN$_2$ we had on-board. To further reduce the thermal background, we cooled the front end of the camera with dry ice beginning at 1800 UT, about 80 minutes before the total eclipse observation.

\begin{figure}[h]
	\centering
	\includegraphics[angle=0,width=1\linewidth]{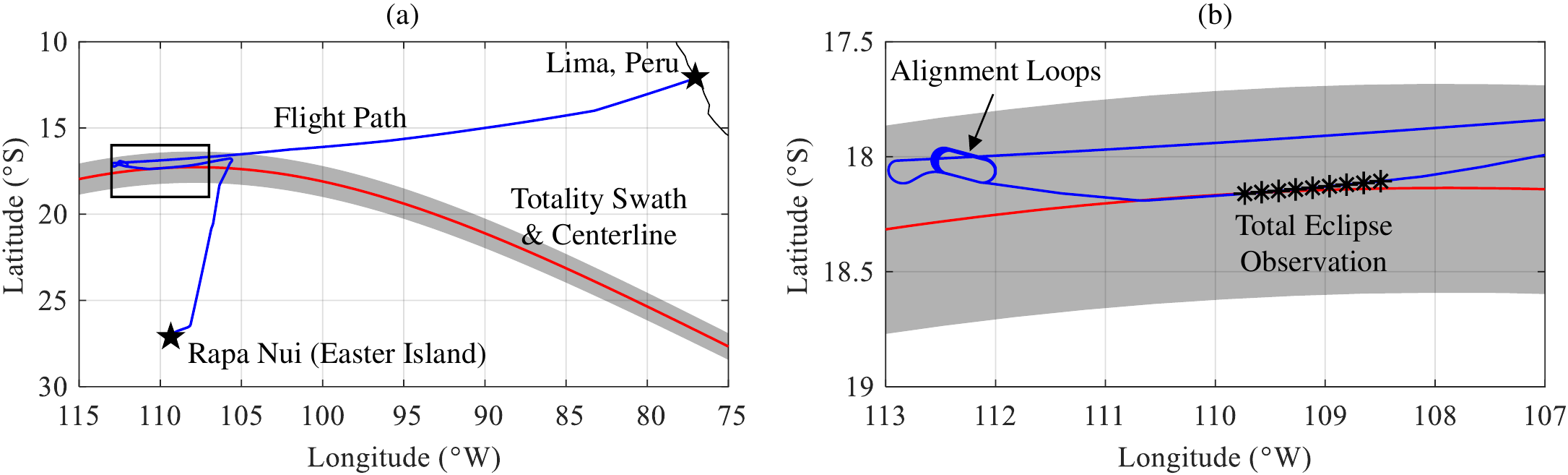}
	\caption{Overview of the eclipse flight (a), with the boxed region enlarged in (b). The ``total eclipse observation'' asterisks mark the location of the GV in one minute increments.}
	\label{fig:flight-map}
\end{figure}

The GV took off from Lima at 1230 UT and flew west for 5.5 hours, entering a holding pattern at about 1800 UT. The holding pattern served two purposes: it allowed us to absorb any schedule delays, and it gave us a number of opportunities to observe the Sun before it was fully eclipsed. The eastward legs of the loops were aligned such that the Sun was observable through the viewport, and we used each eastward pass to fine-tune the periscope and set the spectral alignment. On the last eastward leg, we continued onto the eclipse centerline and waited for the Moon's shadow to overtake us.  The wind, which was 40 m/s out of the west, abbreviated the eastward alignment passes but lengthened the amount of time we spent in totality. Even with the reduced time for alignment, the instrument was ready to observe when the total eclipse began. After the observation, the GV flew south to Rapa Nui (Easter Island), where it refueled for the trip back to Lima. Maps of the flight are shown in Figure \ref{fig:flight-map}.

\subsection{Summary of Observations}

\begin{figure}[h]
	\centering
	\includegraphics[angle=0,width=0.4\linewidth]{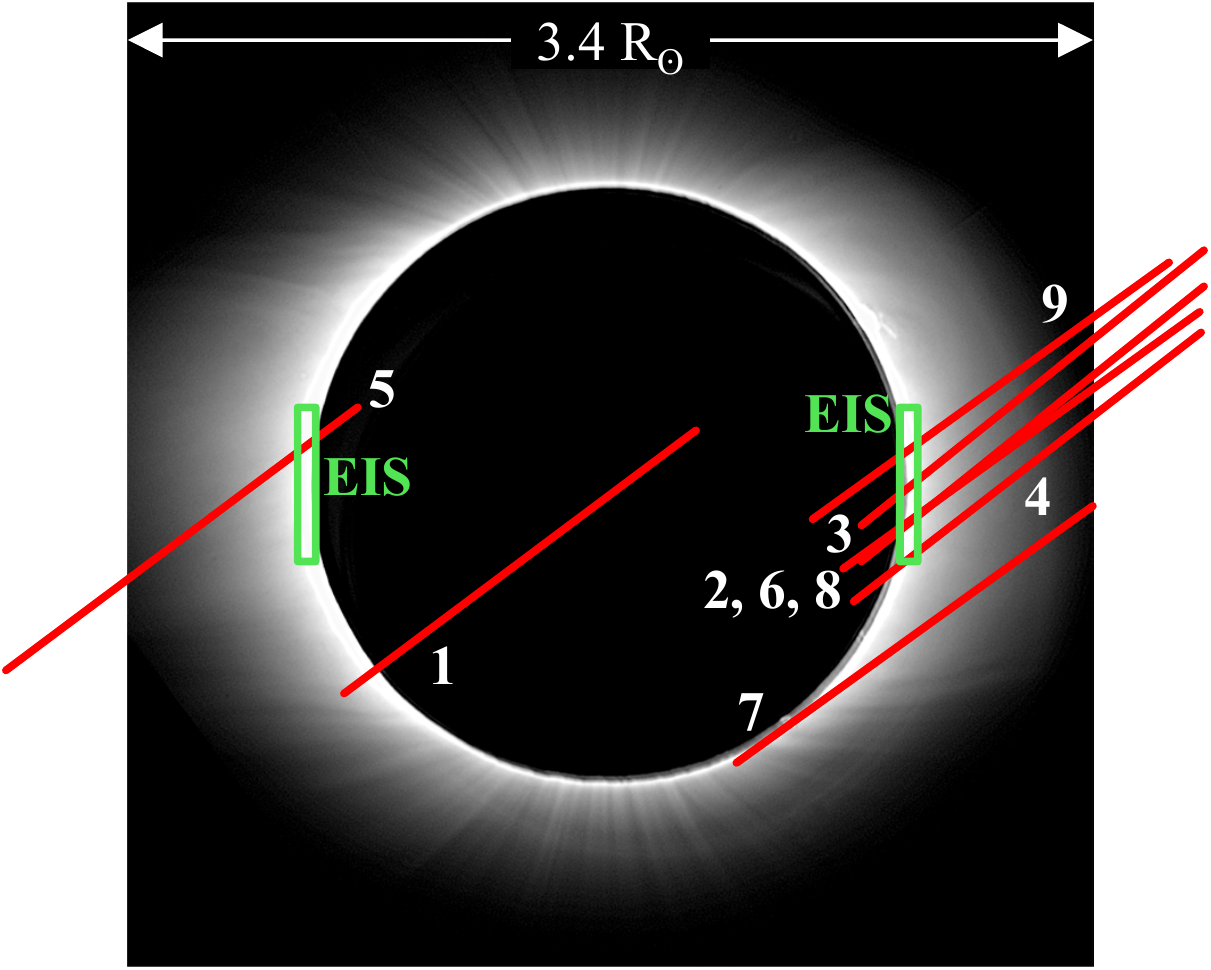}
	\caption{AIR-Spec slit positions superimposed on slit-jaw image.}
	\label{fig:obs-summary}
\end{figure}

Our observation of totality lasted for 9 minutes (19:18:16--19:26:48 UT) and produced 7 minutes of data in 9 slit positions, shown in Figure \ref{fig:obs-summary} and listed in Table \ref{tab:obs-list}. We briefly lost the pointing between positions 4 and 5, but the dark data taken during that period turned out to be invaluable for improving the background subtraction scheme (Section \ref{sec:dark-sub}). Our observations focused on the east and west limbs, where models predicted we would see the brightest intensity and where we had co-located observations with Hinode/EIS. All coronal data (positions 1--8) were taken with an exposure time of 0.25~s. The data shown in this paper come from slit positions 5 and 6, which we refer to as ``east limb'' and ``west limb,'' respectively.
 
\begin{deluxetable}{lcc}
\tablecaption{AIR-Spec slit positions. Data from the highlighted positions are shown in this paper.}
\label{tab:obs-list}
\tablehead{\colhead{Slit Position} & \colhead{Start Time} & \colhead{Duration (s)}} 
\startdata
1. Southeast limb & 19:18:16 & \phn 17.25 \\
2. West limb (EIS) & 19:18:43 & \phn 57.75 \\
3. West limb (EIS) & 19:19:41 & \phn 43.75 \\
4. West limb & 19:20:25 & \phn 53.00 \\
\multicolumn{1}{>{\columncolor{LightCyan}[1.05\tabcolsep]}l}{5. East limb (EIS)} & \multicolumn{1}{>{\columncolor{LightCyan}[1.5\tabcolsep]}c}{19:22:34} & \multicolumn{1}{>{\columncolor{LightCyan}[1.05\tabcolsep]}c}{109.00} \\
\multicolumn{1}{>{\columncolor{LightCyan}[1.05\tabcolsep]}l}{6. West limb (EIS)} & \multicolumn{1}{>{\columncolor{LightCyan}[1.5\tabcolsep]}c}{19:24:38} & \multicolumn{1}{>{\columncolor{LightCyan}[1.05\tabcolsep]}c}{\phn 64.75} \\
7. Prominence & 19:25:58 & \phn 17.50 \\
8. West limb (EIS) & 19:26:16 & \phn 32.00 \\
9. Chromosphere & 19:26:48 & \phn 26.50 \\
\multicolumn{2}{l}{\textbf{{Total Observation Duration:}}} & \textbf{{421 s (7 m)}}  \\
\enddata
\end{deluxetable}

The AIR-Spec observations consist of two-dimensional white-light slit-jaw images and IR spectra imaged along the slit in three passbands near 1.5 \mic, 2 \mic, and 3 \mic. The 3 \mic\ and 1.5 \mic\ spectra are diffracted in first and second order, respectively, and overlap on the top half of the detector.  The 2 \mic\ channel is diffracted in second order and appears alone on the bottom half of the detector.

\section{Spectral Analysis}
\label{sec:analysis}

Fits to the AIR-Spec spectra provide measurements of wavelength and FWHM in all four emission lines as a function of slit position, absolute line radiance as a function of slit position and solar radius, and the relative intensity fall-off of the continuum. The pixel-to-wavelength mapping, instrumental linewidth, and effective area are byproducts of the spectral analysis, and provide a check on the laboratory calibrations. In the next three subsections, we describe how the data were prepared for analysis, outline the routine used for spectral fitting, and report the measured calibration products. The scientific results (line and continuum measurements) are discussed in Section \ref{sec:results}.

\subsection{Data Preparation}

Before beginning the spectral fits, we prepare each exposure by subtracting a dark frame and replacing defective pixels. Each pixel along the slit is indexed to helioprojective coordinates, and all of the frames in a given slit position are co-aligned along the slit. The temporal drift and spatial distortion of the wavelength axis are measured in preparation for producing a unique wavelength calibration at each time step and spatial pixel. 

\subsubsection{Dark Subtraction}
\label{sec:dark-sub}

The thermal instrument background changes over the course of totality as the dry ice sublimates and the front end of the camera warms up. Because of this temporal variation, we cannot subtract darks taken before or after totality from the total eclipse data. Instead, we model the underlying instrument background at the time of each exposure and subtract a modeled dark frame from each frame of data. 

\begin{figure}
\gridline{\fig{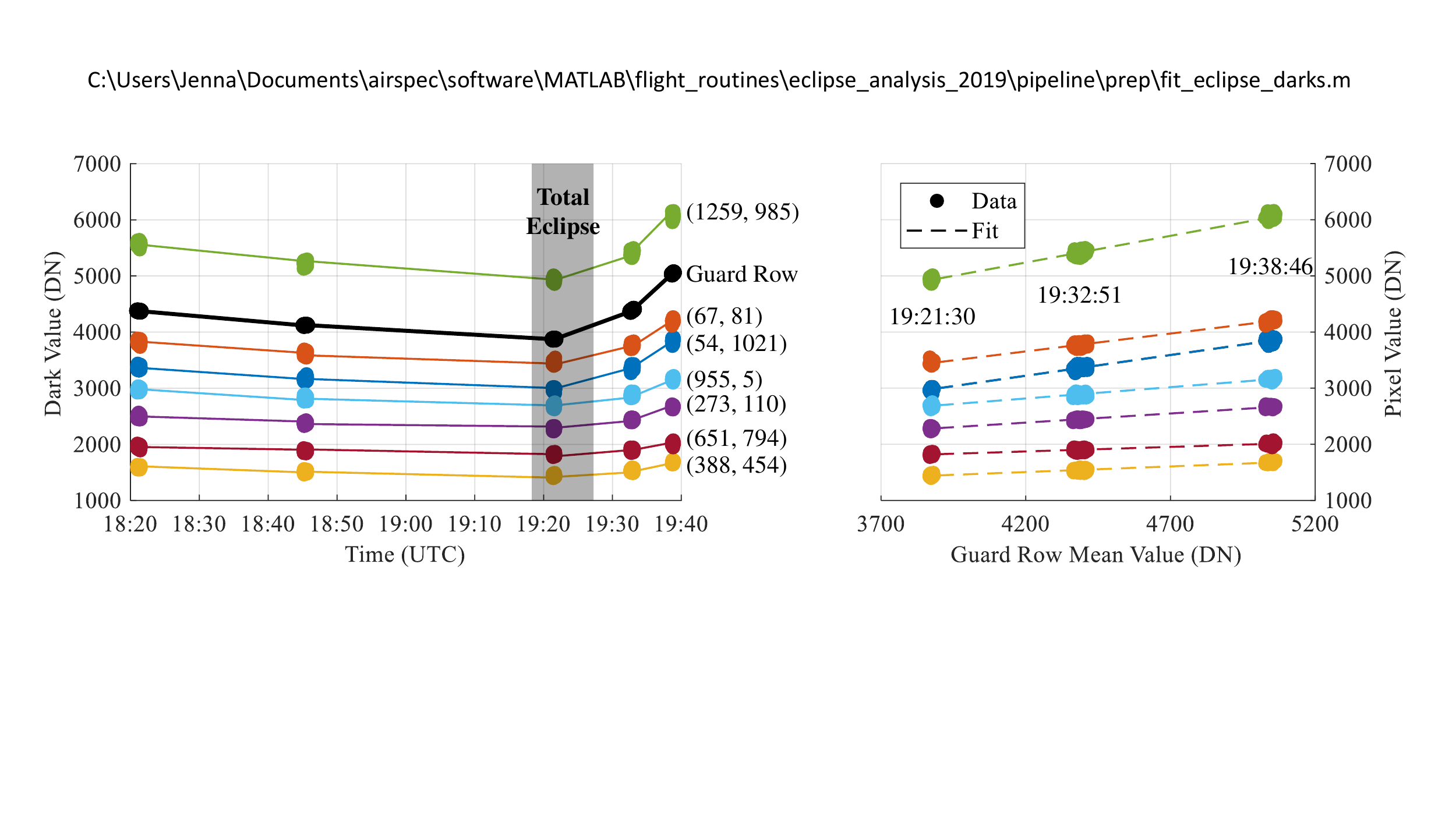}{3.852in}{(a)}
          \fig{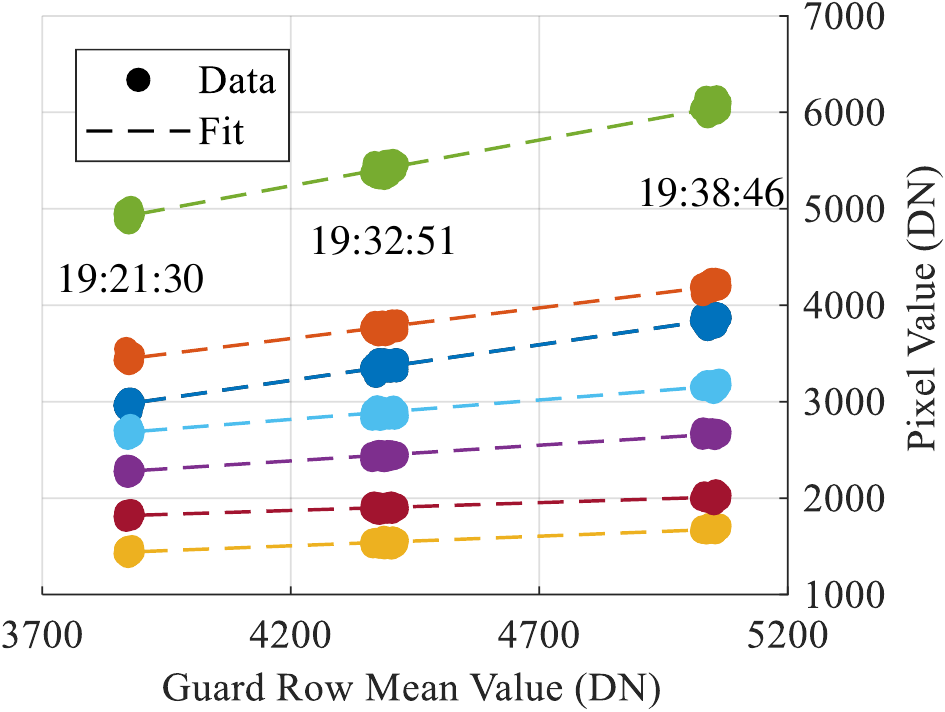}{2.8in}{(b)}}
\gridline{\fig{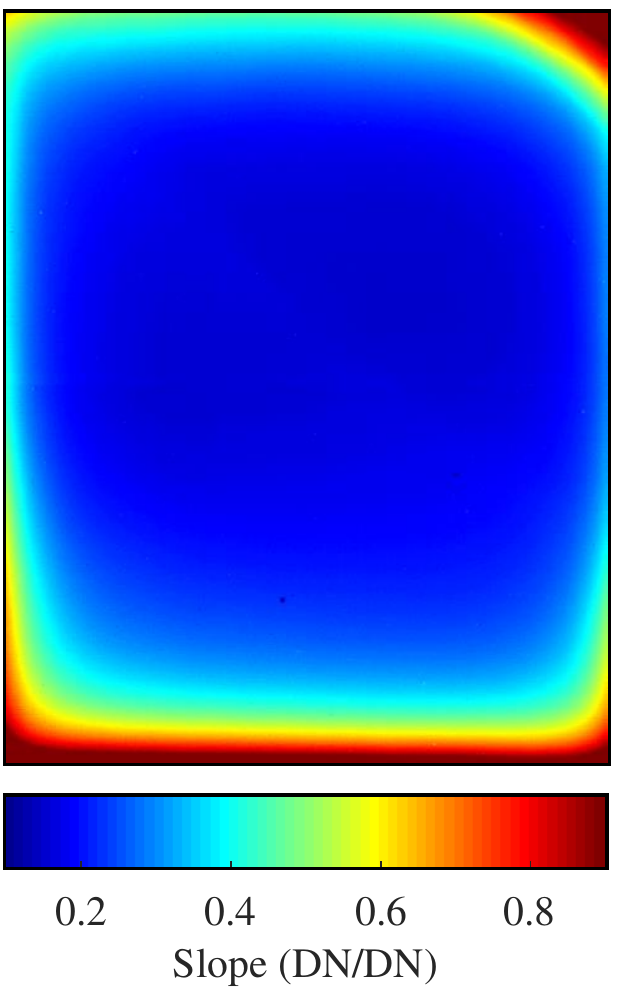}{0.25\linewidth}{(c)}
          \fig{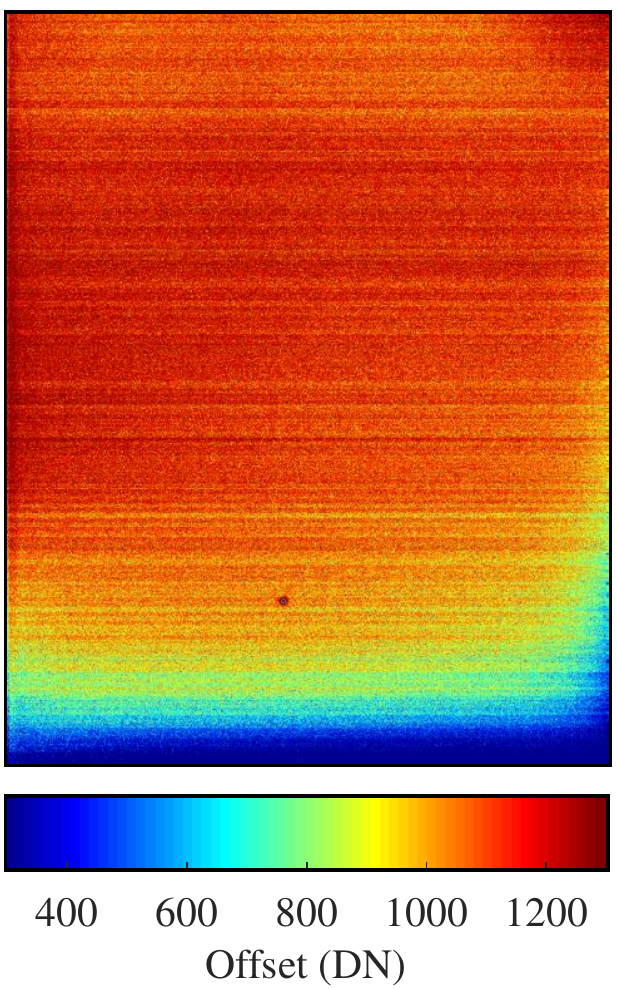}{0.25\linewidth}{(d)}
          \fig{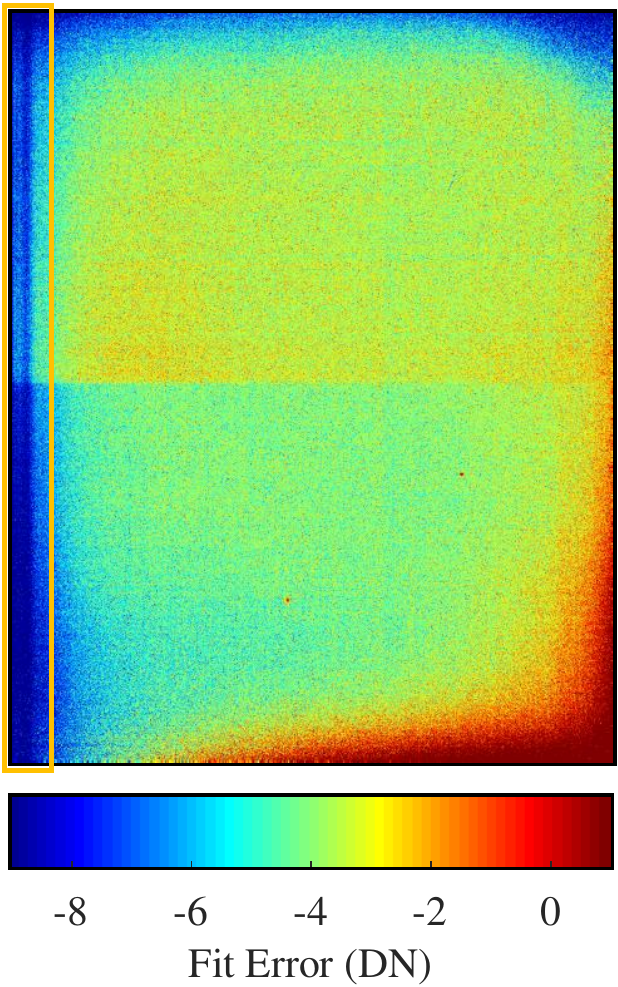}{0.25\linewidth}{(e)}}
\caption{Dark fitting scheme. (a) Dark background as a function of time in seven randomly chosen pixels and the guard row. (b) Dark background as a function of gaurd row mean. The relationship is very close to linear. (c) Slope for the linear mapping from the guard row mean to all pixels. (d) Offset for the linear mapping from the guard row mean to all pixels. (e) Error in dark prediction at 19:21:30 UT. The boxed area highlights a sharp increase in the error near the left edge of the detector.}
\label{fig:dark-sub}
\end{figure}

Figure \ref{fig:dark-sub}(a) shows the dry ice cool-down and warm-up as it appears in seven randomly chosen  pixels and in the whole first row of the detector, which we call the ``guard row.''  The warm-up does not affect the whole detector uniformly; the intensity increases most quickly in pixels near the edge of the detector.   The mean value of the guard row has an approximately linear relationship to the dark value of each pixel, as shown in Figure \ref{fig:dark-sub}(b). The slope and offset that map the guard row to each pixel are shown in Figures \ref{fig:dark-sub}(c) and \ref{fig:dark-sub}(d). 

The guard row gets no light through the optical system because of a slight misalignment between the slit and the detector.  We can therefore use it as a tracer for the dark background over the total eclipse observation. We estimate the dark background $D_{i,j}$ of pixel ($i$, $j$) as

\begin{equation}
    D_{i,j}=b_{i,j}+m_{i,j}\cdot\frac{1}{1024}\sum_{j=1}^{1024}{D_{1,j}}
\end{equation}

\noindent where $i$ is the row (spatial) index, $j$ is the column (spectral) index, $D_{1,j}$ is the dark value in each pixel of the guard row, and $m_{i,j}$, $b_{i,j}$ are the slope and offset that relate each pixel to the the mean of the guard row.  

Figure \ref{fig:dark-sub}(e) shows the error when this relationship is used to predict a set of darks around 19:21:30 UT. Over most of the detector, the error is only a few counts and varies smoothly.  A sharp increase in error (orange box) appears near the left edge of the detector, which is fortunately far from the emission lines.

\subsubsection{Defective Pixel Replacement}

Bad pixels are identified by taking a $5\times5$ median filter of an average dark frame and comparing it with the original frame.  Defective pixels are defined as those further than 500 DN from their median filtered value. This method captures roughly the same pixels as in 2017 \citep[Sec. 3.2.2]{Samra2021}, but the procedure is less subjective. Bad pixels make up 2.3\% of the detector and are replaced using bilinear interpolation.

\subsubsection{Pointing Determination and Spatial Co-Alignment}
\label{sec:coalign}
Next, we map spatial pixel to helioprojective-Cartesian coordinates ($\theta_x,\theta_y$) in each spectrometer frame, and we co-align the IR spectra along the slit. To determine the pointing of each IR exposure, we use context images from the white light slit-jaw camera that provides feedback for the image stabilization system (Section \ref{sec:img-stab}). The camera has a 3.4 \Rs\ field of view (Figure \ref{fig:obs-summary}) and a 3.14 arcsec/pixel plate scale, similar to the 2.31 arcsec/pixel plate scale of the spectrometer. Its 50 Hz frame rate is 12.5x faster than the frame rate of the IR camera, so the slit-jaw frames can be very accurately interpolated to the IR time base. We use the following procedure for pointing determination and spatial co-alignment:

\begin{enumerate}[noitemsep,topsep=0pt]
    \item Rotate each slit-jaw frame to put solar north up.
    \item Fit a circle to the lunar limb to find the pixel that corresponds to Moon center.
    \item Add the angular offset between the Sun and the Moon to find the pixel that corresponds to Sun center.
    \item Align all slit-jaw frames with respect to Sun center.
    \item Map slit-jaw images to solar coordinates using the the 3.14 arcsec/pixel plate scale of the imager.
    \item Find solar coordinates for one end of the slit using the slit-jaw image.
    \item Map every other pixel along the slit to solar coordinates using the 2.31 arcsec/pixel plate scale of the spectrometer. The small pointing variations across spectral pixel are negligible compared to the pointing jitter and are not accounted for.
\end{enumerate}

After finding solar coordinates for each pixel along the slit, we co-align the IR exposures so that the lunar limb appears at the same pixel in every exposure. This overlaps the brightest rows in each frame, providing the highest SNR when averaging the data.

\subsubsection{Measurement of Spectral Drift and Distortion}
\label{sec:spec-drift-dist}

\begin{figure}[h]
	\centering
	\includegraphics[angle=0,scale=0.78]{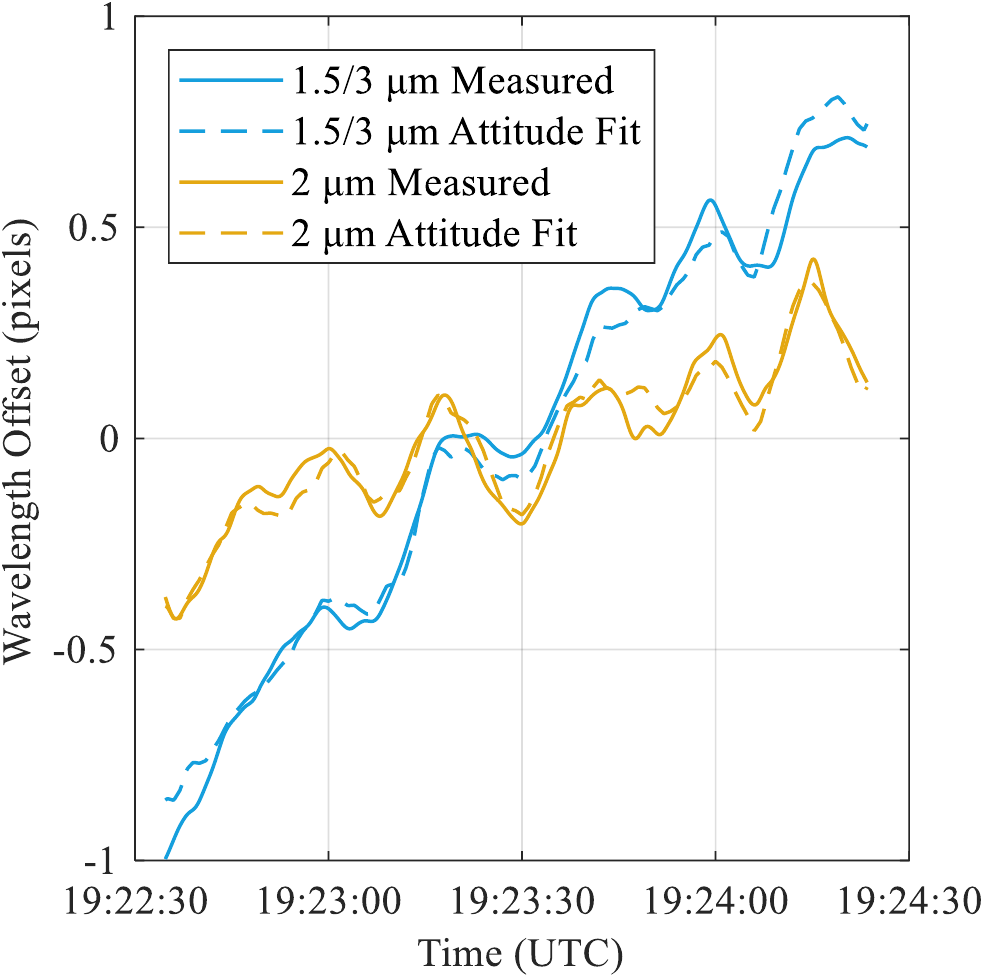}
	\caption{Wavelength offset vs. time for the east limb observation.}
	\label{fig:wave-time}
\end{figure}

Each AIR-Spec channel has a linear mapping from spectral pixel to wavelength. The slope (dispersion) is fixed, but the offset (start wavelength) varies in time and with position along the slit.  \revis{These variations are not intrinsic to the spectrometer design but arise from the fact that it was not implemented to be completely impervious to thermally induced drift and mechanical noise.} Thermal changes in the spectrometer cause a secular drift in the start wavelength, and variations in aircraft attitude result in low-frequency jitter. As a result, the temporal trend is well-described by a linear combination of yaw, pitch, roll, their temporal derivatives, and time (Figure \ref{fig:wave-time}). In 2017, we relied on this type of linear model to calibrate the wavelength offset as a function of time \citep[Sec 3.3.1]{Samra2021}. The SNR of the 2019 data is high enough for us to directly measure the time dependence by cross-correlating a smoothed, spatially averaged spectrum from each frame with the average spectrum over all frames.

\begin{figure}[h]
	\centering
	\includegraphics[angle=0,scale=0.78]{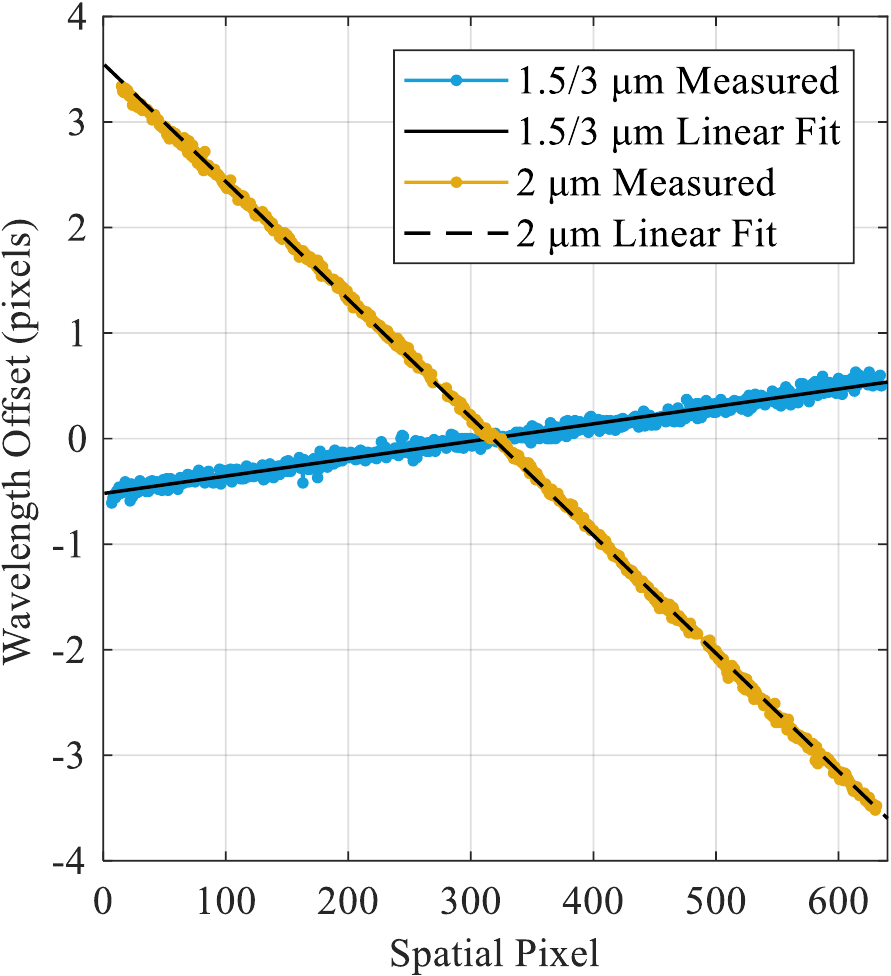}
	\caption{Wavelength offset vs. spatial pixel.}
	\label{fig:wave-spatial}
\end{figure}

The start wavelength varies with position along the slit (spatial pixel) due to distortion in the optical design.  This variation can be mitigated by rotating the slit, but it cannot be nulled out in both channels at the same time.  Instead of applying the geometric correction from 2017 \citep[Sec. 3.2.3]{Samra2021}, a bilinear interpolation which adds some noise, we report the wavelength offset as a function of position along the slit. The offsets, shown in Figure \ref{fig:wave-spatial}, are given by cross correlation of the spectrum in the middle of the slit with the spectrum at every other pixel. The spectra come from a measurement of the uneclipsed solar photosphere taken during a test flight on June 30, 2019. The offset is linear with respect to spatial pixel.

\subsection{Spectral Fitting Routine}
\label{sec:fit-routine}

At a given wavelength $\lambda$, the measured 1.5/3 \mic\  spectrum can be described as

\begin{equation}
    I(\lambda) = ILS(\lambda)\otimes\bigg(\big[K(\lambda)+E(\lambda)\big]\cdot T(\lambda)\cdot A(\lambda) + \big[K(2\lambda)+E(2\lambda)\big]\cdot T(2\lambda)\cdot A(2\lambda)\bigg)
\end{equation}

\noindent where $\lambda$ is the second order wavelength, $I$ is the dark-subtracted detector signal in DN, $ILS$ is the instrument line-spread function, $K$ is the radiance of the coronal continuum in \specrad, $E$ is the radiance of the emission corona in \specrad, $T$ is the transmission through Earth's atmosphere, and $A$ is the product of throughput in cm$^{2}$ sr \AA, exposure time in seconds, quantum efficiency in e$^-$/photon, and detector gain in DN/e$^-$. The second order wavelength $\lambda$ maps to spectral pixel $x$ as

\begin{equation}
\lambda=\lambda_0+\Delta\lambda \cdot x. 
\end{equation}

\noindent The 2 \mic\ channel has the same form, but A(2$\lambda$) = 0 for the whole wavelength range because first order (4 \mic) light is cut off by the focal plane filter. 

With knowledge of $K$ and $T$, we can fit the instrumental parameters $\lambda_0$, $\Delta\lambda$, $ILS$, and $A$ as well as the coronal emission lines $E$.
\revis{For an estimate of the atmospheric transmission, we use the model described in Section \ref{sec:intro} and shown in the right-hand panel of Figure~\ref{fig:atm-model} after convolution with the instrument line-spread function}. As a function of radius $R$, the continuum brightness $B_c$ is modeled as \citep{November1996}

\begin{equation}
    B_c=\left(\frac{0.0551}{R^{2.5}} + \frac{1.939}{R^{7.8}} + \frac{3.670}{R^{18}}\right)\times10^{-6}
    \label{eq:continuum}
\end{equation}

\noindent times the solar brightness \Bs, which we assume to be the radiance of a 5800 K blackbody. 

\begin{figure}[h]
	\centering
	\includegraphics[angle=0,width=0.5\linewidth]{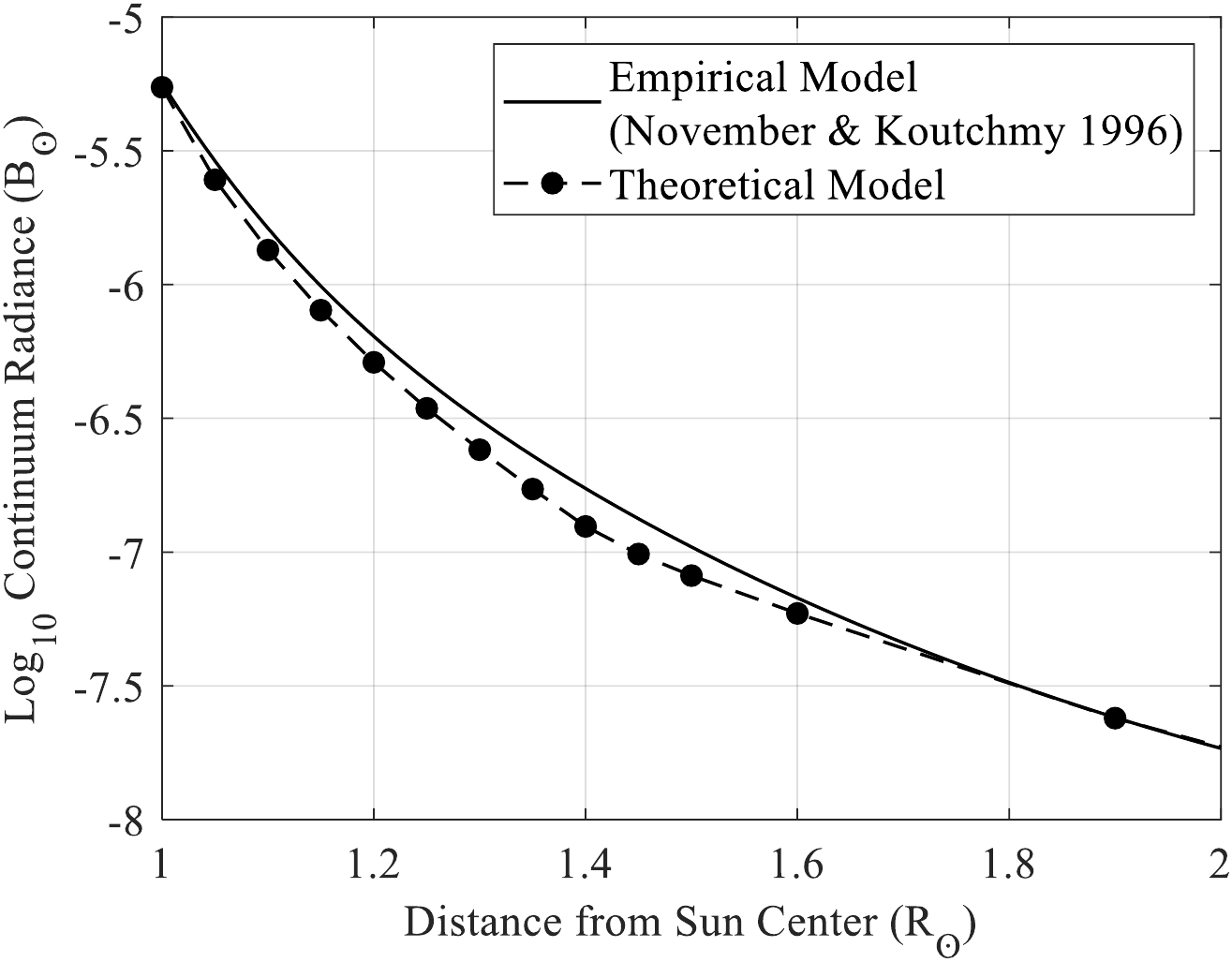}
	\caption{Modeled continuum radiance as a function of radius.}
	\label{fig:cont-model}
\end{figure}

\revis{Equation \ref{eq:continuum} is based on visible-light observations of the corona \citep{November1996}, but we expect the 1.5--3 \mic\ AIR-Spec continuum to have the same intensity in units of \Bs. In the inner corona, the continuum is dominated by wavelength-independent Thomson scattering of photospheric light by free electrons. The scattering cross section does not depend on wavelength and wavelength is not modified by the scattering process. As shown in Figure \ref{fig:cont-model}, the empirical model in Equation \ref{eq:continuum} agrees with a simple wavelength-independent theoretical model that calculates the continuum brightness as the product of scattering cross section, electron density, and dilution factor integrated along the line of sight (G. Del Zanna private communication to J. Samra 2021).	The expected wavelength-independence of the IR inner coronal continuum has also been borne out by  experiment. During the 1994 total eclipse, \cite{Kuhn1996} measured the continuum between 1 and 1.5 \mic\ and found its brightness to be independent of wavelength. An apparent intensity change near 1.2--1.4 \mic\ was attributed to interference from cirrus clouds.}

\begin{figure}[h]
	\centering
	\includegraphics[angle=0,width=0.8\linewidth]{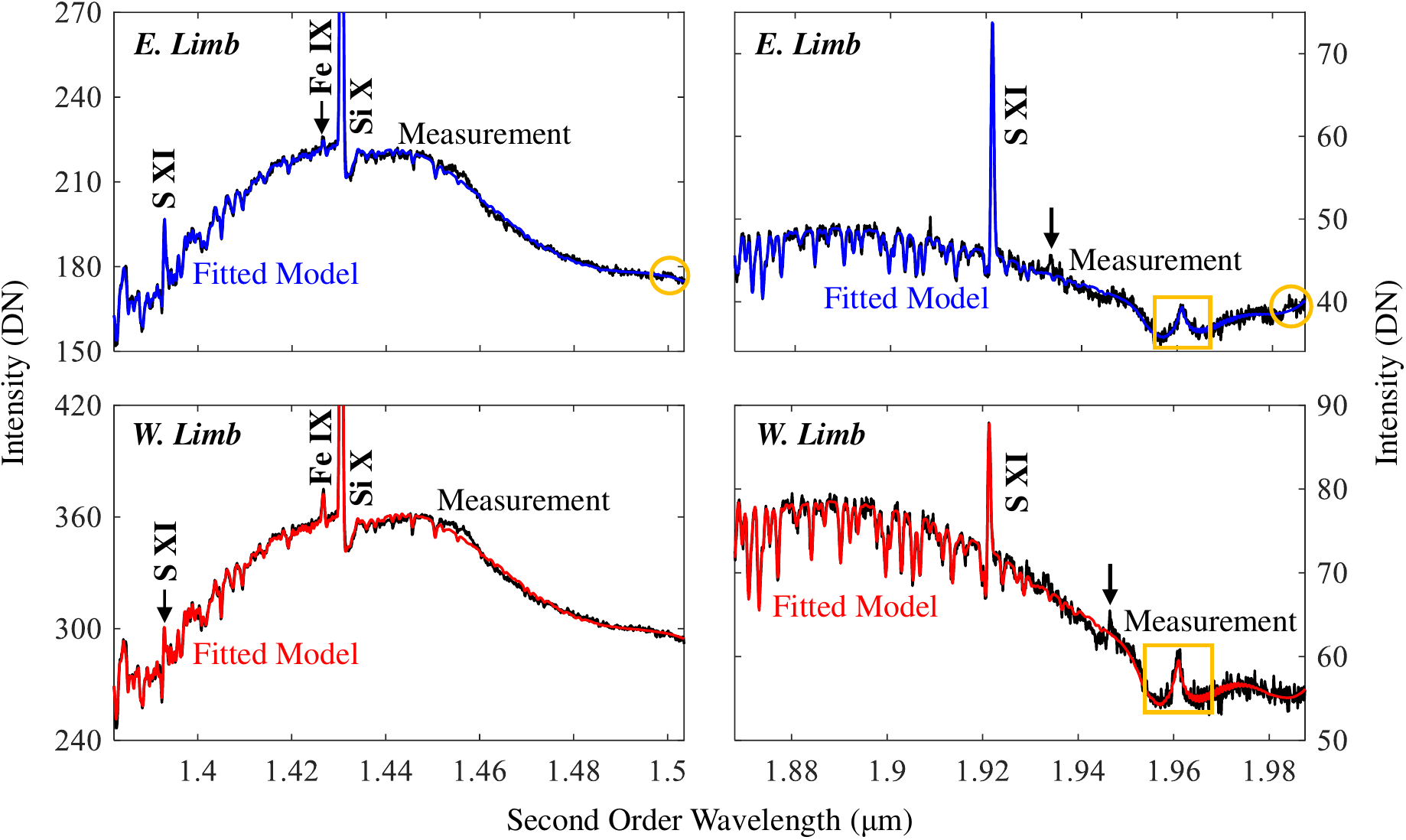}
	\caption{Measured and fitted spectra on the east and west limbs. Errors in the dark prediction are responsible for the circled artifacts in the east limb measurements. \revis{The boxed feature arises from CO$_2$ absorption in Earth's atmosphere. Black arrows mark probable but unconfirmed coronal line detections at 1.934 \mic\ on the east limb and 1.947 \mic\ on the west limb.}}
	\label{fig:spectral-fit}
\end{figure}

In the first stage of our fitting routine, we average each of the data sets in Table \ref{tab:obs-list} in time and over 20 pixels (49~arcsec) along the slit starting at the lunar limb. We mask out the emission lines and use a nonlinear optimization routine to fit $\lambda_0$, $\Delta\lambda$, $ILS$, and $A$, assuming $ILS$ is a Gaussian with unit area and $A$ is a 12\textsuperscript{th} order polynomial (1.5/3~\mic\ channel) or 8\textsuperscript{th} order polynomial (2 \mic\ channel).  This results in the wavelength mapping, instrumental linewidth, and throughput described in Section \ref{sec:cal_products}. Each emission line is given by a Gaussian fit to 

\begin{equation}
    I(\lambda) - ILS(\lambda)\otimes\bigg(K(\lambda)\cdot T(\lambda)\cdot A(\lambda) + K(2\lambda)\cdot T(2\lambda)\cdot A(2\lambda)\bigg)
\end{equation}

\noindent in a 50 \AA\ region centered on the approximate line wavelength.  The data and total fits (continuum and line) are shown in Figure \ref{fig:spectral-fit} for slit positions 5 (east limb) and 6 (west limb).

Next, the line fits are subtracted from the measured spectrum to provide an estimate of the measured background at the line wavelengths. The measured background is used as a basis function in the second stage of the fitting routine. In this stage, we step along the slit in increments of 10 pixels (23 arcsec), averaging 21 pixels (49 arcsec) at each location. With the lines masked out, we find the scale factor that best  fits the background basis function to the measured background data at this new location along the slit. This gives a measure of the reduction in continuum intensity compared to the limb observation. We then subtract the  fitted background  from the data and fit the lines as Gaussians with constant offsets. To account for the gradual variation in instrumental linewidth across the detector, we constrain each FWHM to be $\pm2$\% of the previous estimate.  The second stage of the fitting routine results in line intensities as a function of radius, which are discussed in Section \ref{sec:results}.

\revis{The spectra in Figure \ref{fig:spectral-fit} include several line-like features outside the four confirmed coronal lines. Yellow circles mark the dark prediction artifact highlighted in Figure \ref{fig:dark-sub}e. The boxed 1.961 \mic\ ``line'' is a CO$_2$ absorption feature in Earth's atmosphere. What appears to be emission is actually the gap between the P and R vibrational absorption bands \citep{Atkins2010}. The features at 1.934 \mic\ and 1.947 \mic, marked by black arrows, are probable but unconfirmed  detections of the 2s~2p~$^3$P$^o_{2\rightarrow1}$ transition in \ion{Si}{11} and the 3s$^2$~3p$^4$~3d~$^4$F$_{7/2\rightarrow9/2}$ transition in \ion{Fe}{10}, respectively \citep{Judge1998,DelZanna2018}. These features are near the AIR-Spec detection limit, and further analysis is needed to confirm or deny these identifications and precisely measure the line wavelengths if the detections are confirmed.}

\subsection{Wavelength, Instrumental Linewidth, and Throughput Calibration}
\label{sec:cal_products}

The wavelength mapping, instrumental linewidth, and throughput are byproducts of the fitting routine described in Section \ref{sec:fit-routine}. As an example, Table \ref{tab:wave-fit} reports the start wavelength, dispersion, and instrumental linewidth for the west limb observation in the 21 spatial pixels (49 arcsec) nearest the lunar limb. The values are consistent with lab measurements made with mercury-neon and xenon emission lamps.  By applying the corrections measured in Section \ref{sec:spec-drift-dist}, the start wavelength can be expanded as a function of time and spatial pixel. The dispersion and instrumental linewidth are constant over time and spatial pixel. 

\begin{deluxetable}{cccc}
\tablecaption{Wavelength mapping and instrumental linewidth, west limb.}
\label{tab:wave-fit}
\tablehead{\colhead{Channel} & \colhead{Start Wavelength (\AA)} & \colhead{Dispersion (\AA/pixel)} & \colhead{Inst. FWHM (\AA)}} 
\startdata
1.5 \mic & 15035.06(34)& -1.18607(39)&\phantom{0}5.13(10)\\
\phantom{1.}2 \mic & 19874.98(24) & -1.16732(28)&\phantom{0}6.21(10)\\
\phantom{1.}3 \mic & 30070.12(69) & -2.37215(77)&10.26(21)\\
\enddata
\end{deluxetable}

\revis{The instrumental FWHM are dominated by the slit width of 10 \AA\ in first order (5 \AA\ in second order). The slit width was chosen to maximize throughput at the expense of spectral resolution. As a result, the instrumental linewidths are 2--3 times larger than the expected natural linewidth of the target lines. This makes it impractical to estimate non-thermal velocity from the AIR-Spec measurements, but line intensity and line-of-sight velocities can still be measured precisely. In the emission line FWHM reported in Section \ref{sec:results}, no attempt was made to remove the instrumental FWHM.}

\begin{figure}[h]
	\centering
	\includegraphics[angle=0,width=0.38\linewidth]{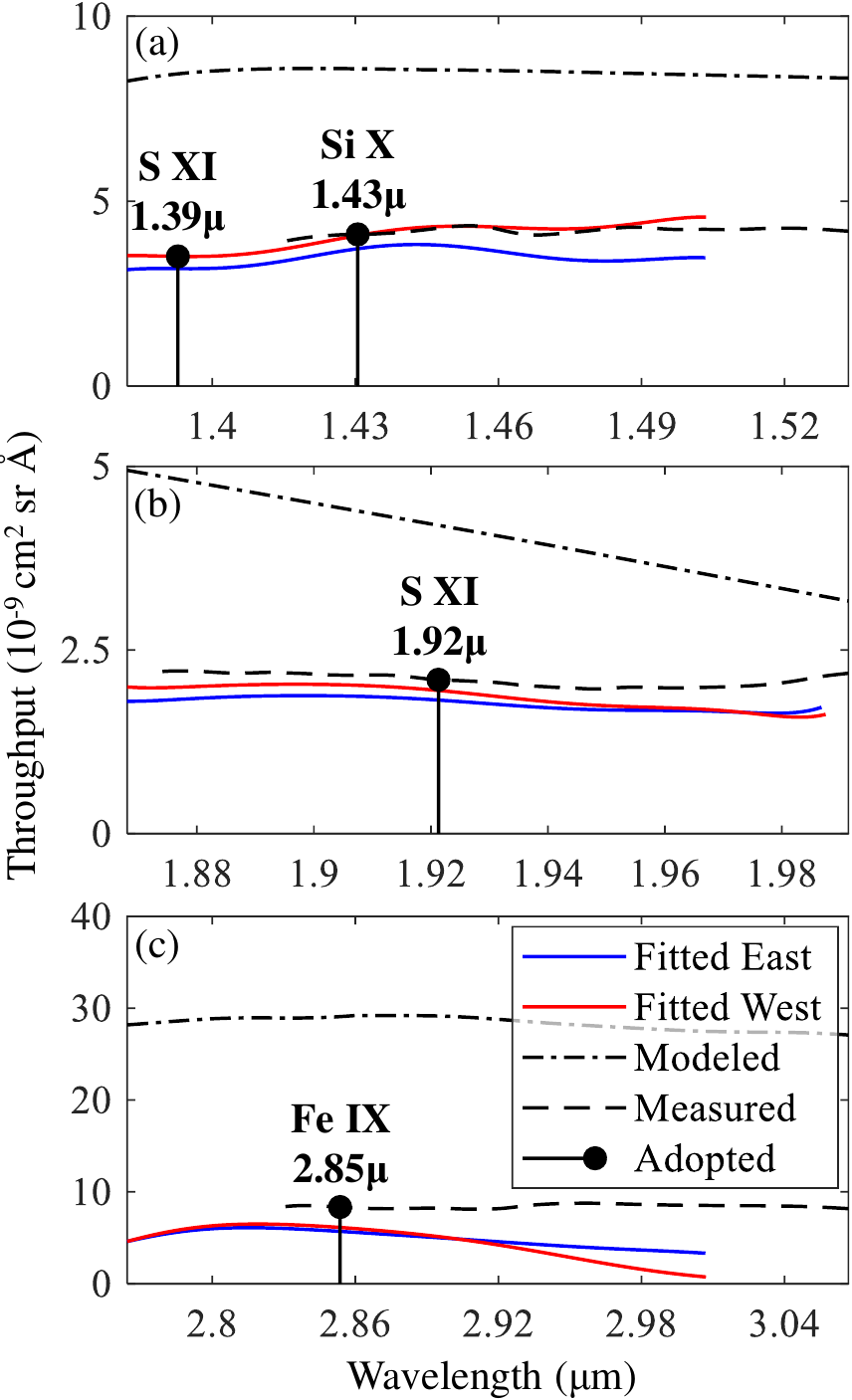}
	\caption{Comparison of modeled, measured, and fitted throughput in the 1.5 \mic\ channel (a), 2 \mic\ channel (b), and 3 \mic\ channel (c).}
	\label{fig:throughput}
\end{figure}

The throughput from the fitting routine is plotted in Figure \ref{fig:throughput} along with the measured and modeled throughput from 2017 \citep[Sec. 3.3.3 and Fig. 23]{Samra2021}.   The fitted throughput agrees very well with the measured throughput. The 2x--3x error in the modeled throughput is unsurprising given that the model makes assumptions on the reflectivity of each protected-silver mirror and transmission through each sapphire window, and we don't precisely know these values or how dust or degradation has changed them.  Due to the large number of surfaces in the light path, a systematic overestimation of even a few percent in the efficiency of each surface will add up to a large overestimate of the overall throughput.

The small disagreements between the fitted and measured throughput are likely due to errors in the dark subtraction and continuum model.  In addition, the 1.5 \mic\ and 3 \mic\ channels are difficult to separate where there is little structure in the atmospheric absorption. In light of these likely errors in the fitted throughput, we adopt the measured throughput at 1.431, 1.921, and 2.853 \mic\ for the analysis in the remainder of this paper. The 2017 spectrum did not extend to 1.391 \mic, so at that wavelength we adopt the fitted throughput on the west limb (position 6), which is very close to the measured at 1.431~\mic. The adopted throughput values are shown in Table \ref{tab:throughput}.

\begin{deluxetable}{ccc}
\tablecaption{Adopted throughput.}
\label{tab:throughput}
\tablehead{\colhead{\multirow{2}{*}{Ion}} & \colhead{Wavelength} & Throughput\\[-5pt]
 & \colhead{(\mic)} & (cm$^2$ sr \AA) }
\startdata
\ion{S}{11} & 1.393& $3.51\times 10^{-9}$\\
\ion{Si}{10} & 1.431& $4.10\times 10^{-9}$\\
\ion{S}{11} & 1.921 & $2.09\times 10^{-9}$\\
\ion{Fe}{9} & 2.853 & $8.35\times 10^{-9}$
\enddata
\end{deluxetable}

\section{Results and Discussion}
\label{sec:results}

\begin{figure}[h]
	\centering
	\includegraphics[angle=0,width=1\linewidth]{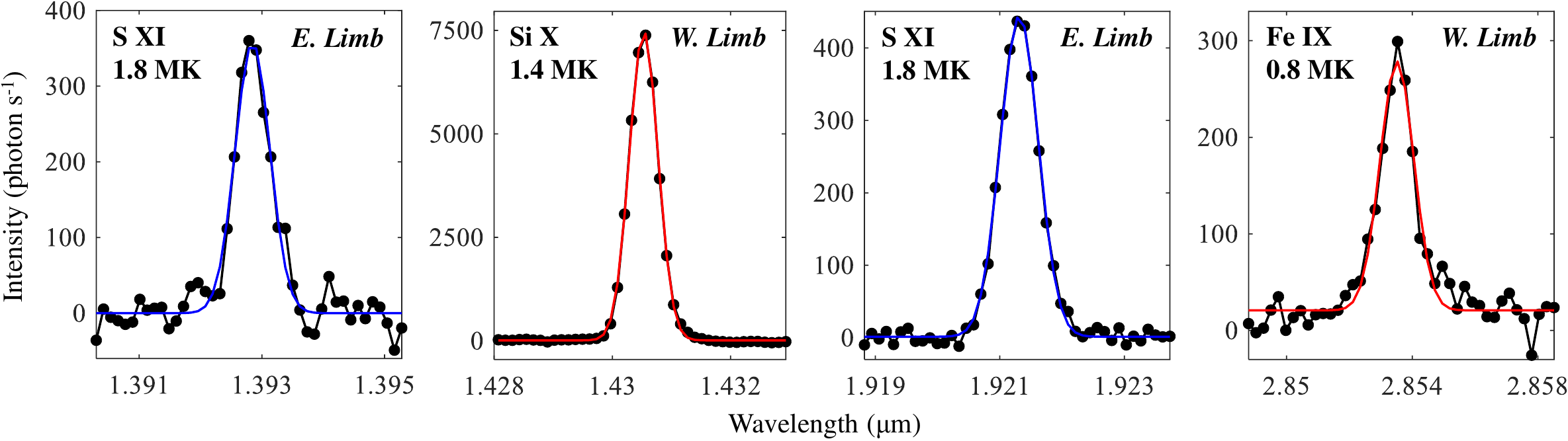}
	\caption{Emission line measurements and Gaussian fits over the 21 pixels (49 arcsec) nearest the lunar limb, corresponding to $R=1.07$ \Rs\ on the east limb and $R=1.04$ \Rs\ on the west limb.}
	\label{fig:line-fits}
\end{figure}

The line and continuum fits described in Section \ref{sec:fit-routine}  provide measurements of center wavelength, FWHM, and absolute intensity for each of the four target lines and a measurement of relative intensity fall-off for the continuum. The line and continuum intensity are functions of solar radius, while the center wavelength and FWHM are reported only for the highest-SNR measurement at the lunar limb. Figure \ref{fig:line-fits} shows the line fits in either the east or west slit position over the 21 pixels (49 arcsec) closest to the lunar limb. The measurement of \ion{S}{11} 1.393 \mic\ is the first of that line \citep{DelZanna2018}.

Table \ref{tab:line-params} lists the wavelengths, intensities at 1.07 \Rs, and FWHM of all four lines in the east and west slit positions.  Wavelength and FWHM uncertainties consider both  the error in the Gaussian fits and the error in the wavelength mapping  (Table \ref{tab:wave-fit}). The FWHM uncertainties are dominated by the line fits, but the wavelength uncertainties have a significant contribution from the wavelength mapping. After solar rotation is corrected, the east and west limb observations have the same line wavelengths to well within the measurement uncertainties. The small FWHM differences are likely due to slight changes in the instrumental linewidth across the detector.  However, intensity differences between the slit positions are significant compared to the uncertainty and cannot be explained by the instrument. We believe they indicate that the corona is significantly hotter near the east limb than the west. \revis{This hypothesis is supported by temperature estimates from coordinated EIS observations and is} the subject of a future paper.

\begin{deluxetable}{ccccc}
\tablecaption{Measured line parameters.}
\label{tab:line-params}
\tablehead{\colhead{Ion} & \colhead{Position} & \colhead{Wavelength (\AA)} & \colhead{Intensity, 1.07 \Rs} & \colhead{FWHM (\AA)}} 
\startdata
\multirow{2}{*}{\ion{S}{11}} & East & 13928.57(62)\phantom{0}& \phantom{0}7.25(32)&\phantom{0}6.67(31)\phantom{0}\\
& West & 13928.60(54)\phantom{0}& \phantom{0}3.93(35)&\phantom{0}6.38(52)\phantom{0}\\
\hline
\multirow{2}{*}{\ion{Si}{10}} & East & 14305.32(51)\phantom{0}& 75.79(51)&\phantom{0}5.98(4)\phantom{00}\\
& West & 14305.21(42)\phantom{0}& 73.39(47)&\phantom{0}5.67(3)\phantom{00}\\
\hline
\multirow{2}{*}{\ion{S}{11}} & East & 19213.27(36)\phantom{0}& 16.25(21)&\phantom{0}7.21(9)\phantom{00}\\
& West & 19213.24(30)\phantom{0}& \phantom{0}7.50(30)&\phantom{0}7.07(21)\phantom{0}\\
\hline
\multirow{2}{*}{\ion{Fe}{9}} & East &28534.75(135)& \phantom{0}1.05(17)&12.77(215)\\
& West & 28534.96(88)\phantom{0}& \phantom{0}2.67(23)&12.67(60)\phantom{0}\\
\hline
\enddata
\tablecomments{Wavelengths are given in vacuo and corrected for solar rotation. \revis{Integrated} intensity is given in units of $10^{11}$~phot~s$^{-1}$~cm$^{-2}$~sr$^{-1}$. FWHM includes the instrumental and natural linewidth.}
\end{deluxetable}

\begin{figure}[h]
	\centering
	\includegraphics[angle=0,width=1\linewidth]{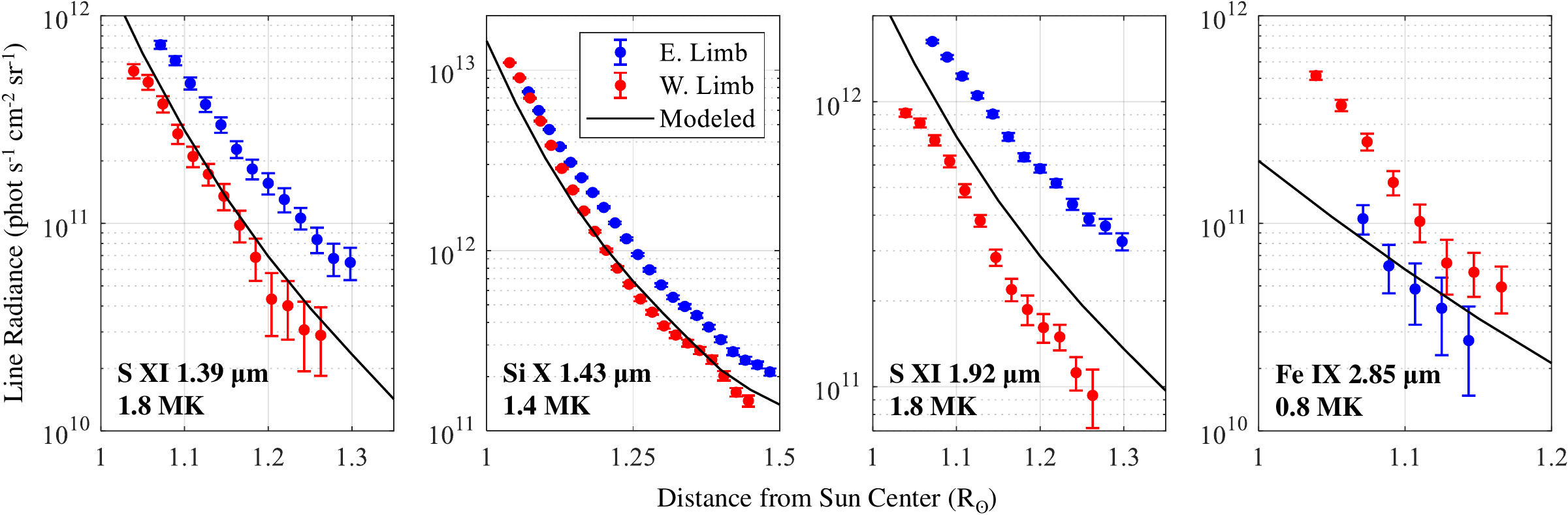}
	\caption{\revis{Integrated} line radiance as a function of radius, measured above the east and west limbs and modeled.}
	\label{fig:intensity}
\end{figure}

Figure \ref{fig:intensity} shows the intensity of the four lines as a function of solar radius in both slit positions. The measurements of intensity fall-off in \ion{S}{11} 1.393~\mic\ and \ion{Fe}{9}~2.853 \mic,  made possible by the AIR-Spec upgrade, are the first for those lines. Figure \ref{fig:intensity} provides additional evidence of a significant temperature difference between the east and west limbs. The cool \ion{Fe}{9} line is stronger on the west limb, while the hotter \ion{S}{11} lines are stronger on the east limb. 

The AIR-Spec intensity measurements are validated by a model from \cite{DelZanna2018Cosie}, which assumes a constant electron temperature of 1.4~MK,  a radial density profile about 30\% lower than what was obtained from polarized brightness measurements of a streamer in 1996, and photospheric abundances. The model includes photoexcitation and integrates along the line of sight, two important considerations for modeling IR lines in the optically thin corona.  The modeled radiances are in excellent agreement with the 2019 AIR-Spec measurements, as shown in Figure \ref{fig:intensity}.

\revis{In addition, we compare our measurement of the 1.431 \mic\ \ion{Si}{10} line to two earlier observations of the same line. In a 1994 eclipse observation, \cite{Kuhn1996} report a line intensity of 3.3~erg~s$^{-1}$~cm$^{-2}$~sr$^{-1}$ ($2.4\times10^{12}$~phot~s$^{-1}$~cm$^{-2}$~sr$^{-1}$) over 43 arcsec near lunar limb, about three times lower than the AIR-Spec \ion{Si}{10} intensity at 1.07 \Rs\ (Table \ref{tab:line-params}). We consider this to be very good agreement given the likely difference in coronal conditions and the fact that the position of the lunar limb in the 1994 eclipse measurement is unknown to the authors of this paper. \cite{PennKuhn1994} report a slit-averaged intensity of $4.5\times10^{-6}$ \Bs\ and a FWHM of 0.289 nm from a 1993 coronagraphic measurement, corresponding to an intensity of about $4.2\times10^{12}$~phot~s$^{-1}$~cm$^{-2}$~sr$^{-1}$ integrated over the line. Near 1.1 \Rs, the intensity is about 50\% higher than the slit average \citep[Figure 3a]{PennKuhn1994}, bringing it into excellent agreement with to the AIR-Spec value of 7.3--7.6 $\times10^{12}$~phot~s$^{-1}$~cm$^{-2}$~sr$^{-1}$ at 1.07 \Rs.}

\begin{figure}[h]
	\centering
	\includegraphics[angle=0,width=0.58\linewidth]{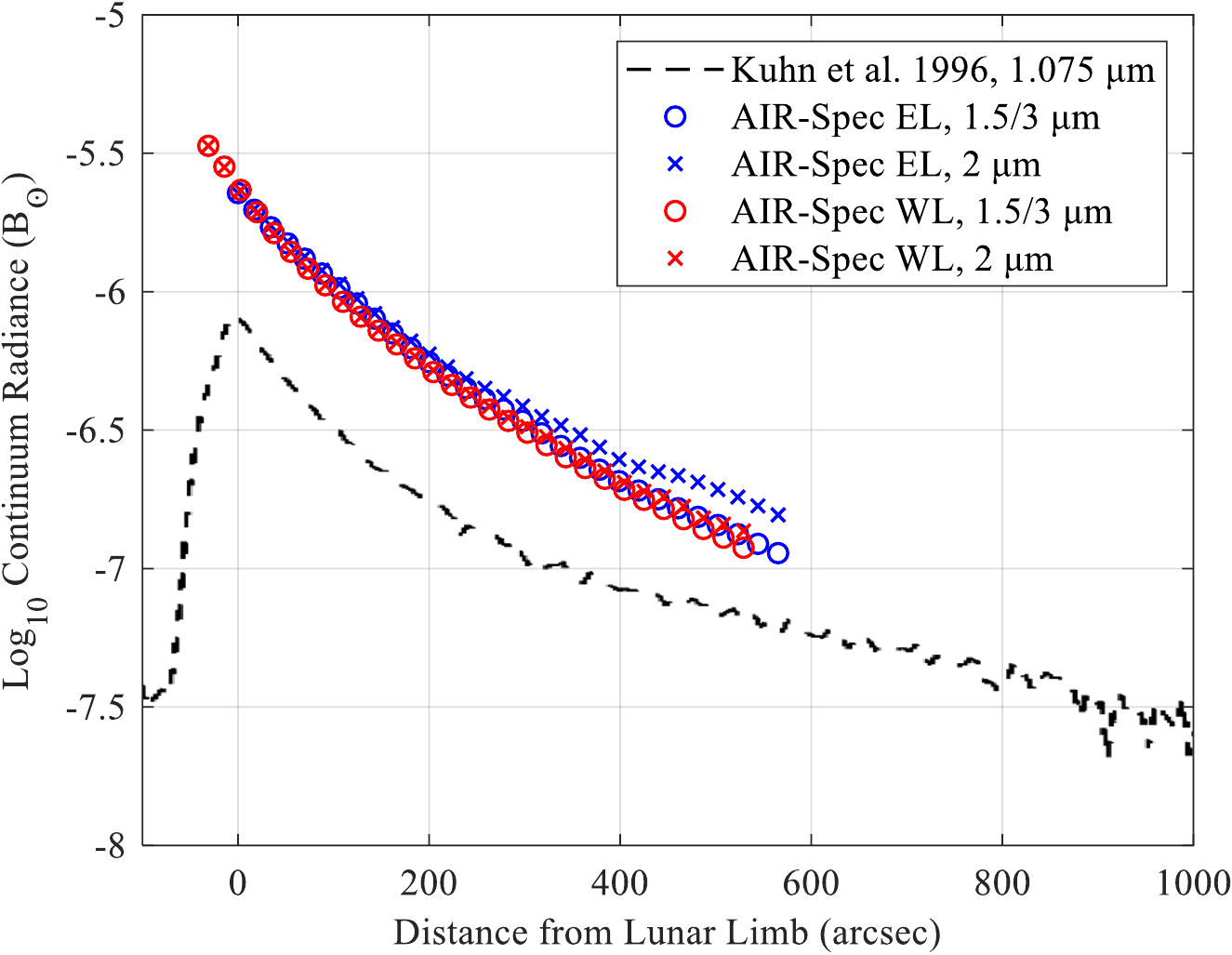}
	\caption{Continuum radiance measured by AIR-Spec and the eclipse experiment described by \citet{Kuhn1996}.}
	\label{fig:cont-meas}
\end{figure}

In Figure \ref{fig:cont-meas}, we compare our measurements of the coronal continuum with \revis{continuum measurements from the same 1994 eclipse experiment. The dashed line, copied from Figure 3b in \cite{Kuhn1996}, shows the 1.075~\mic\ continuum from the earlier experiment. Blue and red symbols show the AIR-Spec measurements on the east and west limbs in the 1.5/3 \mic\ and 2 \mic\ channels. The lunar limb in the 1994 measurement is arbitrarily aligned with 1.07 \Rs, the position of the lunar limb in the eastern AIR-Spec measurement (slit position 5). The AIR-Spec measurements from the overlapped 1.5/3 \mic\ channel agree very well with the 2 \mic\ measurements, suggesting that the continuum brightness is in indeed independent of wavelength as assumed in Section \ref{sec:fit-routine}. The 1994 continuum measurement has approximately the same relative radial fall-off as the AIR-Spec measurements, but its absolute intensity is three times smaller. The same 3x discrepancy was seen in the intensity of the \ion{Si}{10} line, and may be due in part to the different positions of the lunar limb in the two experiments.}

In subsequent flights, AIR-Spec will receive a solar feed from our new large-aperture image stabilization system, the Airborne Stabilized Platform for Infrared Experiments (ASPIRE).  ASPIRE consists of a 30 cm gimballed steering mirror, a gyroscope, and a dedicated high-speed visible light camera. It provides four times the geometric area of the AIR-Spec image stabilization system, improves on the 2019 stabilization algorithm to further reduce drift and jitter, and operates as a self-contained system, allowing it to be used with a variety of focal plane instruments. AIR-Spec and ASPIRE will fly on the GV in the December 4, 2021 eclipse over the Southern Ocean. The new platform enables a 30\% increase in the diameter of the AIR-Spec telescope and makes room for an additional science instrument, a slit-jaw camera that takes two-dimensional images in the 1.4 \mic\ \ion{S}{11} and \ion{Si}{10} lines.

\section{Data Description}
\label{sec:data}
The 2019 AIR-Spec data are available online at the Virtual Solar Observatory. The IR data consist of spatial\;$\times$\;spectral\;$\times$\;temporal cubes in two spectral channels for each slit position. The spectra are in photons/s, but line intensities may be scaled to \specrad\ using the throughput values in Table \ref{tab:throughput}. These data cubes are packaged along with their pointing and wavelength information. Pointing (see Section \ref{sec:coalign}) is a function of spatial pixel and time.   Wavelength (see Sections \ref{sec:spec-drift-dist} and \ref{sec:cal_products}) is a function of spatial pixel, spectral pixel, and time. Co-aligned slit-jaw images are also available.

\acknowledgments 
{The 2019 eclipse observations would not exist without Cory Wolff and the many NCAR Research Aviation Facility technicians, mechanics, pilots, and engineers who made the test flights and eclipse flight possible. We are very grateful to them. We thank SAO engineers Jacob Hohl and Kim Goins for helping us upgrade the instrument and ready it for flight, SAO interns Alisha Vira and Marissa Menzel for their contributions to the closed-loop image stabilization system, \revis{and IRCameras for dramatically reducing the thermal background with their improvements to the camera enclosure.} We thank Giulio Del Zanna for providing the model of the AIR-Spec \revis{and continuum} intensities for inclusion in this paper. We are grateful to Giulio Del Zanna, Helen Mason, and Paul Bryans for their insightful feedback on the 2019 observations and analysis, which greatly improved this paper. The 2019 AIR-Spec upgrade and re-flight was funded by NSF award \#AGS-1822314: \textit{Airborne InfraRed Spectrograph (AIR-Spec) 2019 Eclipse Flight}.
}

\bibliographystyle{aasjournal}
\bibliography{2019_paper.bib}



\end{document}